\documentclass[12pt,a4paper,fleqn]{article}
\usepackage[textheight=23cm,textwidth=16cm]{geometry}

\usepackage{graphicx}
\usepackage{here}
\usepackage{cite}
\usepackage{amsmath}
\usepackage{tabularx}
\usepackage{threeparttable}
\usepackage{amssymb}
\usepackage{setspace}

\newcommand{\onlinecite}[1]{\hspace{-1 ex} \nocite{#1}\citenum{#1}}
\newcommand{\Lower}[1]{\smash{\lower 1.5ex \hbox{#1}}}

\newcolumntype{C}{>{\centering\let\newline\\\arraybackslash\hspace{0pt}}m{0.8cm}}
\newcolumntype{D}[1]{>{\centering\let\newline\\\arraybackslash\hspace{0pt}}m{#1}}
\newcolumntype{E}{>{\centering\let\newline\\\arraybackslash\hspace{0pt}}m{0.9cm}}
\newcommand{\down}{{\bf --} \hskip -0.31cm $\downarrow$ }
\newcommand{\up}{{\bf --} \hskip -0.31cm $\uparrow$ }
\newcommand{\zero}{{\bf --} \hskip -0.31cm \phantom{$\uparrow$} }
\newcommand{\double}{{\bf --} \hskip -0.364cm  $\downarrow \negthickspace \uparrow$ }

\begin{document}

\begin{center}
{\LARGE\bf
Orbital entanglement\\[2ex] in bond-formation processes
}

\vspace{1cm}

{\large
Katharina Boguslawski$^{\rm a}$,
Pawel Tecmer$^{\rm a}$,
Gergely Barcza$^{\rm b}$,
\"Ors Legeza$^{\rm b,}$\footnote{Corresponding Author; E-Mail: olegeza@szfki.hu},
and Markus Reiher$^{\rm a,}$\footnote{Corresponding Author; E-Mail: markus.reiher@phys.chem.ethz.ch}
}\\[4ex]
$^{\rm a}$ETH Zurich, Laboratorium f\"ur Physikalische Chemie, \\
Wolfgang-Pauli-Strasse 10, 8093 Zurich, Switzerland \\
$^{\rm b}$Wigner Research Center for Physics, Strongly Correlated Systems \\
"Lend\"ulet" Research Group, H-1525 Budapest, Hungary\\[2ex]
\end{center}

%\newpage
\begin{abstract}
The accurate calculation of the (differential) correlation energy is central to the quantum
chemical description of bond-formation and bond-dissociation processes.
In order to estimate the quality of single- and multi-reference approaches for this purpose,
various diagnostic tools have been developed. In this work, we elaborate on our previous 
observation [\emph{J. Phys. Chem. Lett.} \textbf{3}, 3129 (2012)] that one- and two-orbital-based 
entanglement measures provide quantitative means for the assessment and classification of 
electron correlation effects among molecular orbitals.
The dissociation behavior of some prototypical diatomic molecules features all types of correlation 
effects relevant for chemical bonding. We demonstrate that our entanglement analysis is convenient to 
dissect these electron correlation effects and to provide a conceptual understanding
of bond-forming and bond-breaking processes from the point of view of quantum information theory.
\end{abstract}
\newpage
%%%%%%%%%%%%%%%%%%%%%%%%%%%%%%%%%%%%%%%%%%%%%%%%%%%%%%%%%%%%%%%%%%%%%%%%%%%%%%%%%%%%%%%%%%%%%%%%%%%%%%%%%%%%%%%%%%%%%%%%%%
\section{Introduction}\label{intro}
The correlation energy is a central quantity in quantum chemistry. It is usually defined as the error
in the electronic energy calculated within the independent-particle model of Hartree--Fock (HF) theory with respect to the exact solution of the 
electronic Schr\"odinger equation \cite{Lowdin_corr_energy,Lowdin_rev},   
\begin{equation}
E^{\mathrm{corr.}}=E^{\mathrm{exact}}-E^{\mathrm{HF}}.
\end{equation}
The exact electronic energy $E^{\mathrm{exact}}$ can be obtained from the full configuration interaction (FCI) approach.
 
Although there exists no rigorous distinction between different types of electron correlation effects,
the correlation energy is typically divided into three categories: \emph{dynamic}, \emph{static}
and \emph{nondynamic} \cite{Bartlett_1994,bartlett_2007}.  
The dynamic correlation energy is considered to be responsible
for keeping electrons apart and is attributed to a large number of configurations (determinants) with
small absolute weights in the wave function expansion, while the nondynamic and static
contributions involve only some determinants with large absolute weights 
which are necessary for an appropriate treatment of the quasi-degeneracy of orbitals
\cite{Sinanoglu1963,Bartlett_1994,bartlett_2007}.
In particular, static electron correlation embraces a suitable combination of determinants
to account for proper spin symmetries and their interactions, 
whereas nondynamic correlation is required to allow a molecule to separate correctly into its
fragments \cite{Bartlett_1994,bartlett_2007}. 

Over the past decades, a number of quantum chemical methods has been developed to accurately
describe either dynamic or nondynamic/static correlation effects.
For instance, M{\o}ller--Plesset perturbation theory and single-reference coupled cluster (CC) theory \cite{bartlett_2007}
are successful in capturing dynamic correlation effects in single-reference cases,
while the complete active space self consistent field (CASSCF) approach \cite{Roos_casscf}
is suitable to describe static correlation in multi-reference problems.
Still, the neglect of one kind of correlation effects may lead to non-negligible errors, which led to the development of "hybrid" approaches,
among which are the complete active space second-order perturbation theory (CASPT2) \cite{caspt21,caspt22,Pulay_CASPT2_2011} 
and the multi-reference CC ansatz \cite{adamowicz-mrcc,Bogus_MRCC,monika_mrcc,Koehn2013}, both with their own intrinsic limitations.   
\textit{A priori} knowledge about the interplay of dynamic, nondynamic and static electron
correlation effects is required to select an appropriate electron correlation method in order to obtain reliable results. 
This issue becomes most severe when spectroscopic accuracy of, say, 0.01 eV for relative energies is desired.
The consideration of dynamic, static and nondynamic correlation effects
on an equal footing still remains a challenge for quantum chemistry.

The quality of single- and multi-reference quantum chemical methods can be estimated by a
number of diagnostic tools \cite{MR_test_Wilson}. Examples are the absolute or
squared weights of the reference or principal configuration (the $|C_0|$ coefficient) obtained
from a CI calculation \cite{lee_1989} and the Euclidean norm of the $t_1$ amplitudes, which are denoted as $T_1$
\cite{Lee_t1_1,lee_1989,T1_open-shell} and $S_2$ diagnostics \cite{Lee_T1_PT} in CC and perturbation theory, respectively. 
Related are the $D_1$ and $D_2$ measures 
based on single and double excitations in single-reference CC theory~\cite{Nielsen1999,Lee_T1_and_D1}.
 
A conceptually different group of diagnostic measures is based on
concepts from quantum information
theory and exploits knowledge about the one-particle reduced density matrix in terms of natural occupation numbers \cite{Ziesche_1995}, 
the two-particle reduced density matrix or its cumulant in terms of the Frobenius norm \cite{Luzanov_2005,Huang2005,Huang2006,Juhasz_2006,Luzanov_2007,Kais2007,
Greenman_2010,Alcoba_2010,Pelzer_2011}, 
the weights from excited configurations of some wave function expansion, \cite{Ivanov_2005} and the distribution of effectively unpaired electrons 
\cite{Takatsuka_1978,Staroverov_2000,Bochicchio_2003,Head-Gordon_2003,Hachmann_2007}.

A complementary classification of electron correlation effects that exploits  entanglement measures among molecular orbitals
was recently proposed by us \cite{entanglement_letter}.
Our analysis is based on the assessment of the entanglement among any pair of orbitals and the entanglement of one orbital
with all other orbitals, respectively, as encoded in a FCI-type wave function.
An in-depth study of iron nitrosyl complexes \cite{entanglement_letter}, featuring complicated electronic structures,  
showed that the static, nondynamic, and dynamic contributions to the correlation energy and,
as a consequence, the  
single- and multi-reference nature of a quantum system, can be distinguished by examining the
entanglement patterns of the orbitals.
Our entanglement analysis comprises two entropic measures: (i) the single-orbital entropy\cite{legeza_dbss},
\begin{equation}
s(1)_i = - \sum_{\alpha} w_{\alpha,i} \ln w_{\alpha,i},
\label{eq:s1}
\end{equation}
($\alpha$ denotes the four different occupations of a spatial orbital)
which quantifies the entanglement between one particular
orbital and the remaining set of orbitals contained in the active orbital space 
from the eigenvalues $w_{\alpha,i}$ of the one-orbital reduced density matrix $\rho_i$
of a given orbital. And, (ii) the mutual information\cite{legeza_dbss,Legeza2006,Rissler2006519}, 
\begin{equation}
I_{i, j} = \frac{1}{2} ( s(2)_{i, j} - s(1)_i - s(1)_j)(1-\delta_{ij}),
\label{eq:I_2}
\end{equation}
which measures the entanglement of two orbitals $i$ and $j$ embedded in the environment of all other active-space orbitals. 
$s(2)_{i,j}$ is the two-orbital entropy between a pair
$(i,j)$ of orbitals, which is calculated from the eigenvalue of the two-orbital reduced density matrix $\rho_{i,j}$ (analogously to Eq.\ (\ref{eq:s1}),
but with $\alpha$ now enumerating the 16 possible no-, one-, and two-electron states defined on the pair of orbitals), and $\delta_{ij}$ is the Kronecker delta. 
The one- and two-orbital density matrices can be determined from a many-particle density matrix by tracing out all many-electron states defined on the active-space orbitals 
that complement those orbital $i$ (and $j$, respectively) whose entanglement with the others shall be studied \cite{Legeza2006}. The general 
occupation-number-vector expansion of the electronic wave function constructed from $L$ active-space orbitals
\begin{equation}
\label{totstatesecqua}
|\Psi \rangle = \sum_{\{n_1\ldots n_L\}} \psi_{n_1,\ldots,n_L}|n_1\ldots n_i \ldots n_j \ldots n_L \rangle,
\end{equation}
($n_i$ denoting the occupation of orbital $i$)
may be decomposed into states $|n_i\rangle$ and $|n_in_j\rangle$ defined on a system comprising the single or pair of orbitals, respectively, and 
into those of its environment defined on the remaining orbitals. If, for the system consisting of one orbital $i$, we split the environment states $e$ ($e$ then being a composite index) as 
$|e_1\rangle \in \{|n_1\ldots n_{i-1}\rangle\}$ and $|e_2\rangle \in \{|n_{i+1}\ldots n_L\rangle\}$,
we can write the total electronic state as
\begin{equation}
|\Psi \rangle ~\rightarrow~ |\Psi^{(n_i,e)} \rangle = \sum_{n_i,e_1,e_2} \psi_{n_i,e_1,e_2} |e_1\rangle \otimes |n_i\rangle \otimes |e_2\rangle .
\end{equation}
For the case of the system consisting of two orbitals $i$ and $j$, we may split the environment states $e$ as
$|e_1\rangle \in \{|n_1\ldots n_{i-1} \rangle\}$, 
$|e_2\rangle \in \{|n_{i+1}\ldots n_{j-1}\rangle\}$, and 
$|e_3\rangle \in \{|n_{j+1} \ldots n_L\rangle\}$. Then, the total electronic state reads in this basis
\begin{equation}
|\Psi \rangle ~\rightarrow~ |\Psi^{(n_i,n_j,e)} \rangle = \sum_{n_i,n_j,e_1,e_2,e_3} \psi_{n_i,n_j,e_1,e_2,e_3} |e_1\rangle  \otimes |n_i\rangle \otimes |e_2\rangle \otimes |n_j\rangle \otimes |e_3\rangle
\end{equation}
The one- and two-orbital density matrix operator can now be expressed as
\begin{equation}
\hat{\rho}_i=\text{Tr}_{e} |\Psi^{(n_i,e)}\rangle  \langle\Psi^{(n_i,e)}| 
\end{equation}
and 
\begin{equation}
\hat{\rho}_{i,j}=\text{Tr}_{e} |\Psi^{(n_i,n_j,e)}\rangle  \langle\Psi^{(n_i,n_j,e)}|,
\end{equation}
respectively. From these operators one may derive the one- and two-orbital density matrices, 
which are calculated from the expansion coefficients $\psi_{n_i,e}$ and $\psi_{n_i,n_j,e}$, respectively.

The total quantum information embedded in a wave function  can be
calculated from the set of single-orbital entropies \cite{legeza_dbss3}
\begin{equation}
I_{\rm tot} = \sum_i s(1)_i.
\label{eq:s_tot}
\end{equation}
As found in Ref.~\citenum{entanglement_letter}, large single-orbital entropies
($s(1)_i> 0.5$) and 
large values for the mutual information $I_{i,j}$ indicate orbitals which are important for nondynamic correlation
effects.
Medium-sized single-orbital entropies ($0.1<s(1)_i < 0.5$) together with moderately entangled orbitals (medium values of $I_{i,j}$ ) 
and small single-orbital entropies ($s(1)_i < 0.1$) accompanied by small 
values of $I_{i,j}$ can be attributed to static and dynamic correlation effects, respectively.
An advantage of our approach is that it reduces electron correlation effects encoded
in a wave function (optimized by any quantum chemical method) to quantities defined for the orbital basis, which can then be
easily compared to each other.

The aim of this paper is
twofold. First, it provides a computational prescription for the calculation of
the entanglement measures introduced in Refs. \citenum{legeza_dbss,Legeza2006,Rissler2006519} and
exploited in Ref.~\citenum{entanglement_letter}, which is given
in section \ref{sec:theory}. 
Second, these entanglement measures are discussed in the context of bond-forming
(or equivalently, bond-breaking) processes in section \ref{sec:bond}
(section \ref{comput_det} contains the computational details), where we demonstrate how the single-orbital entropy
can be employed to monitor the cleavage of chemical bonds. 
Conceptual understanding of electronic structures in terms of
entangled orbitals results and a pictorial representation of how many bonds (single, double,
triple, etc.) are formed between two atoms emerges from the single-orbital entropy diagrams.
We focus on a benchmark set of small diatomic molecules, which constitute
paradigms of both single- and multi-reference problems: the N$_2$, F$_2$ and CsH molecules.

\section{One and two-orbital entanglement measures\label{sec:theory}}

The matrix representation of the one and two-orbital reduced density matrices $\rho_i$ and $\rho_{i,j}$ introduced above can be 
constructed from fermionic correlation functions \cite{Rissler2006519} or from generalized correlation functions\cite{Barcza2013}.
In the following, we present the formalism relying on the latter ones. 
For a spin-1/2 fermionic model, like the one under consideration here, single-electron basis states (orbitals) can be empty,
occupied with an $\alpha$-(spin-up) or a $\beta$-(spin-down) electron,
or doubly occupied with two electrons of paired spin. 
These states we denote as $|$\zero$\rangle$, $|$\down$\rangle$, $|$\up$\rangle$,
and $|$\double$\rangle$, respectively. Since the local basis is four dimensional,
16 possible operators ${\cal O}_i^{(m)}$ arise,
\begin{equation}
{\cal O}_i^{(m)}=\bigotimes_{j=1}^{i-1} \mathbb{I} \otimes {\cal O}^{(m)}  \otimes \bigotimes_{j=i+1}^{L} \mathbb{I}
\end{equation}
which operate on the basis states of a single orbital $i$ 
(with $\mathbb{I}$ being the four-dimensional unit matrix and the action of ${\cal O}^{(m)}$ as summarized in Table \ref{tab:fermion-ops} (for $m=1\dots 16$)).
Note that the dimension of ${\cal O}^{(m)}$ is four, while the (full) operator ${\cal O}_i^{(m)}$ acting on the total state is 4$^L$.

\begin{table}[h!]
\caption{The single-orbital basis operators describing transitions between single-orbital basis states.\label{tab:fermion-ops}}
\centering
\begin{tabular} {c|cccc} 
\hline\hline
${\cal O}^{(m)}$ & \zero &  \down & \up &  \double \\
\hline
\zero &   ${\cal O}^{(1)}$ & ${\cal O}^{(2)}$ & ${\cal O}^{(3)}$ & ${\cal O}^{(4)}$\\
\down &   ${\cal O}^{(5)}$ & ${\cal O}^{(6)}$ & ${\cal O}^{(7)}$ & ${\cal O}^{(8)}$\\
\up &     ${\cal O}^{(9)}$ & ${\cal O}^{(10)}$ & ${\cal O}^{(11)}$ & ${\cal O}^{(12)}$\\
\double & ${\cal O}^{(13)}$ & ${\cal O}^{(14)}$ & ${\cal O}^{(15)}$ & ${\cal O}^{(16)}$\\
\hline\hline
\end{tabular}
\end{table}

The structure of the $4\times 4$ one-orbital operators ${\cal O}^{(m)}$ 
is rather trivial: each operator contains a single element being equal to
one at matrix position $k$, $l$ where $|l\rangle$ is the initial state and $|k\rangle$ is the final state. It thus acts
like a transition matrix from state $|l\rangle$ to $|k\rangle$. Explicitly, we may write the one-orbital 
operator ${\cal O}^{(m)}$ as 
\begin{equation}
\label{fulloneorbop}
{\cal O}^{(1)}=\left(\begin{array}{cccc}
1&0&0&0\\
0&0&0&0\\
0&0&0&0\\
0&0&0&0
\end{array}\right)
~~\mbox{,}~~
{\cal O}^{(2)}=\left(\begin{array}{cccc}
0&1&0&0\\
0&0&0&0\\
0&0&0&0\\
0&0&0&0
\end{array}\right) ~~,
~~\mbox{etc.}~~
\end{equation}
such that the matrix elements of ${\cal O}^{(m)}$ can be expressed as a Kronecker delta, 
$({\cal O}^{(m)})_{k,l}=\delta_{(l+4[k-1]),m}$ for $m=1\dots 16$, and the one-orbital states are  
labeled by $k=1 \dots 4$ and $l=1 \dots 4$.
For the many-electron wave function in occupation-number-vector representation the 
multi-site reduced density matrices can be expressed using products of ${\cal O}_i^{(m)}$ operators 
acting on specific sites.

The one-orbital reduced density matrix $\rho_i$ can then be calculated by taking those operators which
do not change the single-orbital basis state \cite{Rissler2006519}, 
\begin{equation}
\rho_i = \begin{pmatrix}
\langle{\cal O}^{(1)}_i\rangle & 0 & 0 & 0 \\
0 & \langle{\cal O}^{(6)}_i\rangle & 0 & 0 \\
0 & 0 & \langle{\cal O}^{(11)}_i\rangle & 0 \\
0 & 0 & 0 & \langle{\cal O}^{(16)}_i\rangle
\end{pmatrix},
\end{equation}
where the expectation value is calculated from the total electronic state.
We may abbreviate this structure in tabular form as summarized in Table \ref{tab:rho1-fermion}. 
Once the one-orbital reduced density matrix $\rho_i$ is constructed, $s(1)_i$ can be determined from its
eigenvalues $w_{i,\alpha}$ according to Eq.\ (\ref{eq:s1}). 
\renewcommand*{\arraystretch}{1}
\begin{table}
\caption{Expressing  
the single-orbital reduced density matrix, $\rho_{i}$, in terms of single-orbital operators, ${\cal O}^{(m)}_i$, defined in Table \ref{tab:fermion-ops}.
For better readability only the operator number indices, $m$ are shown corresponding to
$\left\langle \Psi\right| O^{(m)}_{i} \left|\Psi\right\rangle $.
}
\begin{tabular}{c|cccc} 
\hline
\hline
 $\rho_{i}$ & \zero &  \down & \up &  \double \\
\hline
\zero & 1 & & &\\
\down &  & 6 & &\\
\up &  &  & 11  &\\
\double &  &  &  & 16 \\
\hline
\hline
\end{tabular}
\label{tab:rho1-fermion}
\end{table}

In the case of the two-orbital reduced density matrix $\rho_{i,j}$, $(\rho_{i,j})_{kl,pq}$ mediates a transition from state $|p,q\rangle$ to $|k,l\rangle$ where $|p\rangle$ and $|k\rangle$ are
the initial and final states, respectively, defined on spatial orbital $i$, while
$|q\rangle$ and $|l\rangle$ are the initial and final states, respectively, defined on spatial orbital $j$.
$\rho_{i,j}$ can be calculated from expectation values of operator products ${\cal O}^{(m)}_i{\cal O}^{(n)}_j$, where $m=p+4(k-1)$ and $n=q+4(l-1)$.
Thus, the two four-dimensional spaces for states defined on orbitals $i$ and $j$ are expressed as one 
16-dimensional space whose basis is labelled $|$\zero \!\zero$\rangle$, $|$\zero \!\down$\rangle$,  $|$\down \!\zero$\rangle$, $|$\zero \!\up$\rangle$, ~\dots, 
$|$\double \!\double$\rangle$. 
As in the one-orbital case, the two-orbital reduced density matrix $\rho_{i,j}$ can be 
built explicitly using the expectation values of the two-orbital correlation functions.
The two-orbital reduced density matrix $\rho_{i,j}$ has non-zero matrix elements only between two-orbital states possessing the same
quantum numbers, $n$ and $s_z$, of two orbitals since $\rho_{i,j}$ does not change the quantum numbers of the two orbitals. 
Therefore, $\rho_{i,j}$ has a block diagonal structure and there is no need to calculate all 
$16\times16$ matrix elements. Taking also into account that $\rho_{i,j}$ is symmetric,
only 26 expectation values remain to be determined.
The single-orbital operator combinations for
orbitals $i$ and $j$ used for obtaining the nonzero matrix elements of $\rho_{i,j}$
are summarized in Table \ref{tab:fermion-ops-2}.
For better readability, we abbreviated $\left\langle \Psi\right| {\cal O}^{(m)}_{i}{\cal O}^{(n)}_{j} \left|\Psi\right\rangle $ by $m/n$ in that Table,
where ${\cal O}^{(n)}_{i}$ is the $n$-th one-orbital operator acting on orbital $i$ as given in Table \ref{tab:fermion-ops-1}
and $|\Psi\rangle$ is again a general correlated wave function.
Although the calculation of the two-orbital correlation functions is expensive since
all $ {\cal O}^{(m)}_{i}{\cal O}^{(n)}_{j}$ terms must be renormalized and stored independently, the
required 26 calculations can be performed in a fully parallel manner \cite{Barcza2013}.
Once the two-orbital reduced density matrix is constructed,
$s(2)_{i,j}$ can be determined from its eigenvalues in analogy to Eq.\ (\ref{eq:s1}), and thereby, the mutual information for
each orbital pair $(i,j)$ can be evaluated.
An important feature of this method is that one can also
analyze the sources of entanglement encoded in $I_{i,j}$ by studying the individual correlation functions \cite{Barcza2013}.

\begin{table}[h!]
\caption{Expressing the two-orbital reduced density matrix, $\rho_{i,j}$, in terms of single-orbital operators, ${\cal O}^{(m)}_i$.
For better readability only the operator number indices, $m$ are shown, thus 
$m/n$ corresponds to  $\left\langle \Psi\right| {\cal O}^{(m)}_{i}{\cal O}^{(n)}_{j} \left|\Psi\right\rangle $.
$n$ and $s_z$, denote the quantum numbers of two orbitals.\label{tab:fermion-ops-2}}
\centering
\scalebox{0.8}{
%\begin{tabular} {c||c|cc|cc|c|cccc|c|cc|cc|c}
\begin{tabular} {C||C|CC|CC|C|CCCC|C|CC|CC|C}
\hline\hline
& \multicolumn{1}{D{0.9cm}|}{{\small $n$=$0$, $s_z$=$0$}} & \multicolumn{2}{D{1.8cm}|}{{\small $n$=$1$, $s_z$=-$\frac{1}{2}$}}
& \multicolumn{2}{D{1.8cm}|}{{\small$n$=$1$, $s_z$=$\frac{1}{2}$}} & \multicolumn{1}{D{0.9cm}|}{{\small $n$=$2$, $s_z$=-$1$}} 
& \multicolumn{4}{D{3.6cm}|}{{\small$n$=$2$, $s_z$=$0$}} & \multicolumn{1}{D{0.9cm}|}{{\small $n$=$2$, $s_z$=$1$}} 
& \multicolumn{2}{D{1.8cm}|}{{\small$n$=$3$, $s_z$=-$\frac{1}{2}$}} & \multicolumn{2}{D{1.8cm}|}{{\small$n$=$3$, $s_z$=$\frac{1}{2}$}}&
\multicolumn{1}{D{0.9cm}}{{\small $n$=$4$, $s_z$=$0$}}\\
\cline{2-17}
$\rho_{i,j}$ & \zero \!\zero & \zero \!\down & \down \!\zero & \zero \!\up &   \up \!\zero & \down \!\down &
 \zero \!\double & \down \!\up & \up \!\down & \double \!\zero 
 & \up \!\up &  \down \!\double  & \double \!\down  &\up \!\double &\double \!\up & \double \!\double\\
\hline
\hline
\zero \!\zero     & 1/1 & & & & & & & & & & & & & &  & \\
\hline
\zero \!\down     &  & 1/6 & 2/5 & & & & & & & & & & & & &\\
\down \!\zero     &  & 5/2 & 6/1 & & & & & & & & & & & & &\\
\hline
\zero \!\up       &  &     &  & 1/11 & 3/9 & & & & & & & & & & &\\
\up \!\zero       &  &     &  & 9/3  & 1/11 & & & & & & & & & & &\\
\hline
\down \!\down     &  &     &  &      &   & 6/6 & & & & & & & & & &\\
\hline
\zero \!\double   &  &     &  &      &   &  & 1/16 & 2/15 & 3/14 & 4/13 & & & & & &\\
\down \!\up       &  &     &  &      &   &  & 5/12 & 6/11 & 7/10 & 8/9 & & & & & &\\
\up \!\down       &  &     &  &      &   &  & 9/8  & 10/7 & 11/6 & 12/5 & & & & & &\\ 
\double \!\zero   &  &     &  &      &   &  & 13/4 & 14/3 & 15/2 & 16/1 & & & & & &\\
\hline
\up \!\up         &  &     &  &      &   &  &      &      &      &   & 11/11 & & & & &\\
\hline
\down \!\double   &  &     &  &      &   &  &      &      &      &   &   & 6/16 & 8/14 & & &\\
\double \!\down   &  &     &  &      &   &  &      &      &      &   &   & 14/8 & 16/6 & & &\\
\hline
\up \!\double     &  &     &  &      &   &  &      &      &      &   &   &      &   & 11/16 & 15/12 &\\
\double \!\up     &  &     &  &      &   &  &      &      &      &   &   &      &   & 15/12 & 16/11 &\\
\hline
\double \!\double &  &     &  &      &   &  &      &      &      &   &   &      &   &       &   & 16/16\\
\hline\hline
\end{tabular}
}
\end{table}

It is worth to note that if the Hilbert space of the wave function is partitioned into a system and an environment part
in a way that both blocks are built up from continuous segments of orbitals, i.e., 
orbitals are permuted so that the orbitals $i$ and $j$ are situated next to each other in the system block (taking care of the proper phase factor
which is then introduced), then Eq.\ (\ref{totstatesecqua}) reduces to the form 
\begin{equation}
\left|\Psi\right\rangle = \sum_{s,e} \psi_{s,e} \left|s\right\rangle \otimes \left|e\right\rangle \;,
\label{eq:bipartite-wavefunction}
\end{equation}
where $\left|s\right\rangle$ stands for the basis states of the system
and  $\left|e\right\rangle$ for those of the remaining orbitals. 
In this representation, the components ${\cal O}^{(m)}$ can be written in terms of spin-dependent, 
fermionic creation $c^\dagger_\sigma$ and annihilation $c_\sigma$
operators, which create and annihilate an electron of $\sigma$ spin,
and spin-dependent number operators $n_{\sigma}$, defined as
\begin{equation}
n_{\sigma} = c^\dagger_\sigma c_\sigma.
\end{equation}
All 16 components ${\cal O}^{(m)}$ of the one-orbital operators are collected in Table \ref{tab:fermion-ops-1}.
As a consequence, the elements of the reduced density matrices can also be expressed using
standard one-orbital operators in this bipartite representation \cite{Rissler2006519}.
We should emphasize that the one-particle reduced density matrix, and hence its
eigenvalue spectrum, contributes only one ingredient of the total 26 orbital
correlation functions (see Table \ref{tab:fermion-ops-1}). Therefore, the one- and two-orbital reduced
density matrices can comprise more information about quantum entanglement and electron
correlation than encoded in the occupation numbers of the one-particle reduced density matrix.
\begin{table}[h!]
\caption{The sixteen possible transitions between the different single-orbital basis states mediated by single-orbital operators.\label{tab:fermion-ops-1}}
\begin{center}
\begin{tabular}{c|c}
\hline\hline
%notation & elementary operators\\
%\hline 
${\cal O}^{(1)}$& $1-n_{\uparrow}-n_{\downarrow}+n_{\uparrow}n_{\downarrow}$\\
${\cal O}^{(2)}$ & $c_{\downarrow}-n_{\uparrow}c_{\downarrow}$\\
${\cal O}^{(3)}$ & $c_{\uparrow}-n_{\downarrow}c_{\uparrow}$\\
${\cal O}^{(4)}$ & $c_{\downarrow}c_{\uparrow}$\\
${\cal O}^{(5)}$ & $c_{\downarrow}^{\dagger}-n_{\uparrow}c_{\downarrow}^{\dagger}$\\
${\cal O}^{(6)}$ & $n_{\downarrow}-n_{\uparrow}n_{\downarrow}$\\
${\cal O}^{(7)}$ & $c_{\downarrow}^{\dagger}c_{\uparrow}$\\
${\cal O}^{(8)}$ & $-n_{\downarrow}c_{\uparrow}$\\
${\cal O}^{(9)}$ & $c_{\uparrow}^{\dagger}-n_{\downarrow}c_{\uparrow}^{\dagger}$\\
${\cal O}^{(10)}$ & $c_{\downarrow}c_{\uparrow}^{\dagger}$\\
${\cal O}^{(11)}$ & $n_{\uparrow}-n_{\uparrow}n_{\downarrow}$\\
${\cal O}^{(12)}$ & $n_{\uparrow}c_{\downarrow}$\\
${\cal O}^{(13)}$ & $c_{\downarrow}^{\dagger}c_{\uparrow}^{\dagger}$\\
${\cal O}^{(14)}$ & $-n_{\downarrow}c_{\uparrow}^{\dagger}$\\
${\cal O}^{(15)}$ & $n_{\uparrow}c_{\downarrow}^{\dagger}$\\
${\cal O}^{(16)}$ & $n_{\uparrow}n_{\downarrow}$\\
\hline\hline
\end{tabular}
\end{center}
\end{table}

In this work, the entanglement measures are determined from wave functions optimized by the
density matrix renormalization group (DMRG) \cite{scholl05,ors_springer,marti2010b,chanreview}
algorithm developed by White \cite{white} since DMRG allows for a balanced description of
nondynamic, static and dynamic electron correlation
effects within a sufficiently large active space. However, we should note that all entropic quantities could also be determined from any
other correlated wave function.
If the DMRG algorithm is used to optimize the electronic wave function,
the one-orbital and
the two-orbital correlation functions can be calculated for all orbitals $i$ and all orbital
pairs $(i,j)$ at the end of a full DMRG sweep.
We should note that all orbital-entanglement functions are determined from well-converged DMRG
wave functions and thus the choice of the DMRG parameter set, e.g., the ordering
of molecular orbitals and the number of renormalized active-system states
(\emph{cf.} section \ref{comput_det}), does not affect the entanglement measures.
The analysis of the one- and two-orbital correlation functions provides a different perspective on some well-known correlation problems.

\section{Computational details}\label{comput_det}
All calculated quantities (energies and entanglement measures) are in Hartree atomic units.

\subsection{Basis sets and relativity}
For the light elements H, N and F, Dunning's aug-cc-pVTZ basis set was used with
the following contractions H: (6$s$3$p$2$d$)$\rightarrow[$4$s$3$p$2$d]$;
N and F: (11$s$6$p$3$d$2$f$)$\rightarrow[$5$s$4$p$3$d$2$f]$).
For the Cs atom, a contracted QZP ANO-RCC basis set
((26$s$22$p$15$d$4$f$2$g$)$\rightarrow[$9$s$8$p$7$d$3$f$2$g]$) was
employed \cite{ANO-RCC_Cs}, which is specifically optimized for the Douglas--Kroll--Hess (DKH)
Hamiltonian \cite{DKH2,Wolf_2002}. Scalar relativistic effects were considered in the case of CsH through the DKH Hamiltonian (tenth order for CASSCF and third order for CC calculations, respectively)
\cite{Reiher_2004a,Reiher_2004b}.
Higher-order DKH Hamiltonians are not implemented in the NWChem 6.1 release and could therefore not be used in the CC calculations. 

\subsection{CASSCF}\label{subsec:casscf}
All CASSCF calculations \cite{Siegbahn_casscf,Roos_casscf} have been performed with the \textsc{Molpro 2010.1} program package \cite{Knowles_1985,Werner_1985,molpro}. 
For the N$_2$ and F$_2$ molecules, all active spaces contained both non-bonding 2$s$-orbitals, the bonding 2$p_{\pi}$ (doubly degenerate) and 2$p_{\sigma}$,
and the antibonding 2$p_{\pi^*}$ (doubly degenerate) and 2$p_{\sigma^*}$ combinations imposing
D$_{2h}$ point group symmetry (\textit{cf.} Table \ref{tbl:resolution}). This corresponds to active spaces comprising ten electrons in
eight orbitals for N$_2$ (CAS(10,8)SCF) and 14 electrons in eight orbitals for F$_2$ (CAS(14,8)SCF).
For the CsH molecule, a CAS(10,15)SCF was employed imposing C$_{2v}$ point group symmetry
(\textit{cf.} Table \ref{tbl:resolution}),
which includes the bonding and antibonding combinations of the Cs 6$s$-orbital
and of the H 1$s$-orbital, respectively, as well as the remaining  nonbonding 4$d$-, 6$p$-, 5$d$-
and 7$s$-orbitals of the Cs atom. Note that the 4$d_{\sigma}$-orbital was not included in the active space in analogy to the work presented in Ref.\ \onlinecite{Reiher_CsH}. 
The missing dynamic correlation energy was added on top of the CASSCF wave functions
employing second order perturbation theory (CASPT2) as implemented in the
\textsc{Molpro 2010.1} program package \cite{CASPT2_molpro}.

\begin{table}[h!] 
     \caption{Resolution of the relevant irreducible representations of D$_{\infty h}$ point group towards those of the  D$_{2h}$ and  C$_{2v}$ subgroups\cite {Altmann}.}
     \label{tbl:resolution}
\begin{tabular}{cccccc}
\hline
\hline
D$_{\infty h}$  && D$_{2h}$&&  C$_{2v}$\\
\hline
$\sigma_g$ & &a$_{g}$                 && a$_{1}$\\
$\sigma_u$  &&b$_{1u}$                 &&a$_{1}$\\
$\pi_g$ && b$_{2g} \oplus$ b$_{3g}$   && b$_{1} \oplus$ b$_{2}$\\
$\pi_u$ && b$_{2u} \oplus $ b$_{3u}$  && b$_{1} \oplus$ b$_{2}$\\
$\delta_g$& & a$_{g} \oplus$ b$_{1g}$ && a$_{1} \oplus$ a$_{2}$\\
$\delta_u$& & a$_{u} \oplus$ b$_{1u}$ && a$_{1} \oplus$ a$_{2}$\\
$\phi_g$ & &b$_{2g} \oplus$ b$_{3g}$  && b$_{1} \oplus$ b$_{2}$\\
$\phi_u$ & &b$_{2u} \oplus $ b$_{3u}$ && b$_{1} \oplus$ b$_{2}$\\
\hline
\hline
\end{tabular}
\end{table}

\subsection{DMRG}
All DMRG calculations were performed with the \textsc{Budapest DMRG} program \cite{dmrg_ors}. As orbital
basis, the natural orbitals obtained from the CASSCF calculations as described in
section \ref{subsec:casscf} are taken. 
The active spaces could be extended to CAS(10,46), CAS(14,32) and CAS(10,51)
in our DMRG calculations for the N$_2$, F$_2$ and CsH molecules, respectively. 
For N$_2$, additionally the 5$\times \sigma_g$, 5$\times \sigma_u$, 4$\times \pi_u$, 4$\times
\pi_g$, 2$\times \delta_u$, 2$\times \delta_g$, 1$\times \phi_g$ and 1$\times \phi_u$ 
lowest lying virtual orbitals have been included in the active space, while for F$_2$, the
4$\times \sigma_g$, 4$\times \sigma_u$, 3$\times \pi_u$, 3$\times \pi_g$, 1$\times \delta_u$
and 1$\times \delta_u$ virtual orbitals have been added to the CAS(14,8) active space
(note that the $\pi$, $\delta$, and $\phi$ are doubly degenerate, see Table \ref{tbl:resolution}).
In the case of the CsH molecule, the CAS(10,15) active space was extended by the
10$\times \sigma$, 6$\times \pi$, 5$\times \delta$ and 2 $\times \phi$ virtual orbitals.    
We should note that for the CsH molecule
at an internuclear distance of 5.5 \AA{} only 50 orbitals were in the active space 
since a $\delta$-type virtual orbital has been rotated into a $\sigma$-type virtual orbital
in the CAS(10,15)SCF reference. Therefore, only 9 out of 10 $\sigma$-orbitals were selected
for the DMRG active space for this particular bond length. 

To enhance DMRG convergence, the orbital ordering was optimized \cite{orbitalordering} and the number of renormalized active-system states was chosen
dynamically according to a predefined threshold value for the quantum information loss
\cite{legeza_dbss} employing the dynamic block state selection approach \cite{legeza_dbss2,legeza_dbss3}. As initial guess, the
dynamically-extended-active-space procedure was applied \cite{legeza_dbss}.
In the DMRG calculations, the minimum and maximum number of renormalized
active-system states $m$ was varied from 512 to 1024 and from 1024 to 2048, respectively,
while the quantum information loss was set to $10^{-5}$ in all calculations.
Note that a large number of renormalized active-system states $m_{\rm start} = 1024$
was chosen for the
initialization procedure for the F$_2$ molecule to achieve fast and stable convergence.
The convergence behavior of all DMRG calculations with respect to the DMRG parameter set is
summarized in the Supporting Information.

\subsection{Coupled cluster}
The (restricted) coupled cluster singles and doubles (CCSD), coupled cluster singles, doubles and perturbative
triples (CCSD(T)) and coupled cluster singles, doubles and triples (CCSDT) 
calculations were performed with the \textsc{NWChem 6.1} quantum chemical program package
\cite{nwchem,nwchem_11,nwchem_web} using the tensor contraction engine 
\cite{tce_1,tce_2,tce_3,tce_4}.
All CC calculations were carried out for two different sizes of the orbital space:
for (i) the complete virtual orbital space (later denoted as full-virtual and labeled as CC($x$,all),
where $x$ indicates the number of correlated electrons)
and (ii) for a (restricted) orbital space of the same size as in our DMRG
calculations (in terms of number of correlated electrons and orbitals). 
The latter was performed to estimate the amount of the dynamic correlation energy
captured within the DMRG active space.
Note, however, that the orbitals in the CC and
DMRG calculations are different because the Hartree--Fock determinant is the reference for the CC calculations.
Hence, the correlation energies obtained from DMRG and CC in the DMRG active space are not identical. Still,
a comparison of the restricted-space CC results with the full CC results, both calculated for Hartree--Fock
orbitals, shows the precentage of electron correlation captured within the active space.

\section{Results and discussion\label{sec:bond}}
Following the introduction of an entanglement classification of correlation effects in Ref.\ \onlinecite{entanglement_letter}, in this section we discuss how the entanglement
measures can be instrumental for an analysis of bond-breaking and bond-forming processes.
It is important to understand that these measures allow us to extract orbital-related information from a correlated wave function.
The single-orbital entropies and mutual information
corresponding to the molecular orbitals forming a chemical bond show large values when
bonds are stretched.  All other orbitals remain slightly entangled with small values for
$s(1)_i$ and $I_{i,j}$. Such patterns are consistent with the
understanding of nondynamic correlation effects, where those orbitals become strongly entangled
which allow a molecule to correctly separate into its fragments. The one- and
two-orbital entanglement measures should, therefore, provide a qualitative
picture of how many bonds are formed between two atoms. 
A similar analysis holds for strongly correlated systems in condensed matter physics when the
strength of entanglement bonds is determined \cite{Barcza2013}.
A qualitative, entanglement-based
bond order can be determined from the total number of steep changes in the $s(1)_i$-diagram
present in the dissociation limit (divided by two to account for the bonding
and anti-bonding combination of molecular orbitals), which will be demonstrated in the following sections. 

Furthermore, the process of bond-breaking or bond-forming along a reaction coordinate can be
monitored in the evolution of the single-orbital entropies. Since static and nondynamic electron
correlation effects become dominant if bonds are stretched, the
single-orbital entropies corresponding to the bonding and antibonding pair of molecular
orbitals should increase gradually. In particular, the rate of growth should
depend on the type (or strength) of a specific bond as the magnitude of the one- and
two-orbital entanglement measures is connected to the structure of the electronic wave function
(\textit{cf.} section \ref{sec:theory}). In a qualitative picture, the $s(1)_i$
values of orbitals involved in weak $\pi$-bonds increase faster than those corresponding
to strong $\sigma$-bonds. A chemical bond is considered broken if the $s(1)_i$ remain unchanged
when the two centers $A$ and $B$ are further pulled apart, \emph{i.e.}, if
${\partial s(1)_i}/{\partial r_{AB}} \rightarrow 0$, and thus
${\partial I_{\rm tot}}/{\partial r_{AB}} \rightarrow 0$ for large bond lengths $r_{AB}$. This should allow us to resolve
bond-breaking processes of individual $\sigma$-, $\pi$-, or $\delta$-bonds in multi-bonded
centers.

In the following, we perform an entanglement analysis of the triple bond in N$_2$ and of the single bonds
in F$_2$ and CsH at various internuclear distances. For this, we present entanglement diagrams that depict the mutual information defined 
among each pair of orbitals. These diagrams are color coded: blue lines indicate a mutual information
whose order of magnitude is 0.1, red lines one of magnitude 0.01 and green lines one of magnitude 0.001.
The color coding is as in Ref.\  \onlinecite{entanglement_letter},
where weakly entangled orbitals important for dynamic correlation effects are connected by green lines, while
orbitals important for static and non-dynamic correlation effects are connected by red and blue lines, respectively.
In addition, each entanglement diagram is accompanied by a single-orbital entropy plot, in which each natural orbital
is assigned the calculated single-orbital entropy.
During the dissociation process, orbitals that are involved in describing the chemical bond will become strongly entangled, i.e., they exhibit
increasing $I_{i,j}$ and single-orbital entropies.

\subsection{The dinitrogen molecule} 

First, we investigate the dissociation process of the dinitrogen molecule.
It is well-known that the restricted Hartree--Fock wave function dissociates into an unphysical
mixture of neutral and ionic fragments with charges varying from $\pm 1$ to $\pm 3$. 
This suggests that a large amount of correlation of different types (nondynamic, static,
and dynamic) is mandatory to properly describe
the dissociation process and spin-recoupling of the triple bond in the N$_2$ molecule
\cite{MRCI_N2,Mitrushenkov_N2,Chan_N2,Shepard_2006,Paldus_N2,Shepard_2008,MRCC_N2_2009}.       

\begin{figure}[H]
\centering
\includegraphics[width=0.9\linewidth]{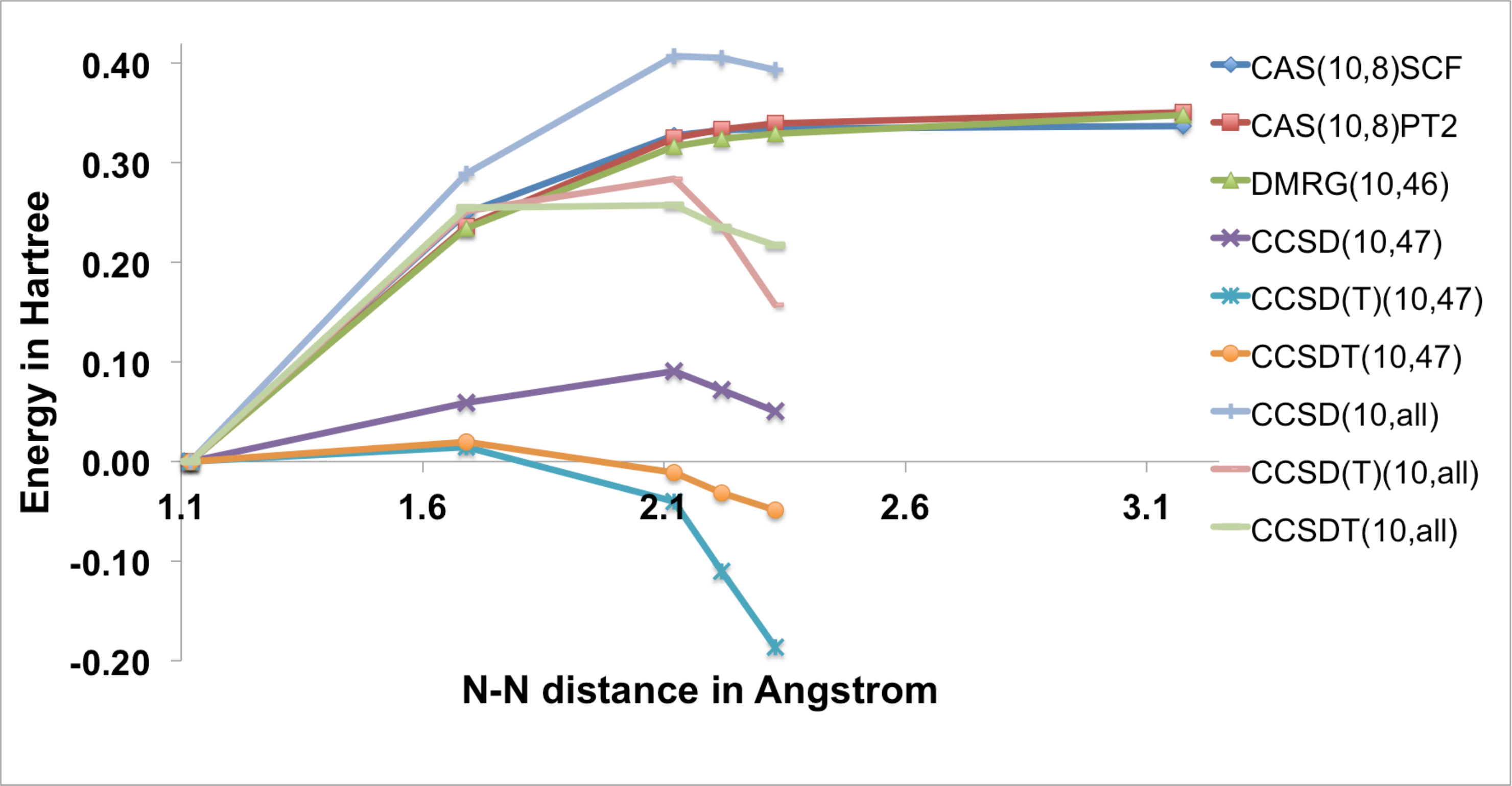}
\caption{
Electronic energy differences for the N$_2$ molecule at various intermolecular distances determined
by different quantum chemical methods. The energy reference is the electronic energy at equilibrium distance of each method.
}\label{fig:n2-energies}
\end{figure}

For the entanglement study, we chose six points along the reaction coordinate, including the
equilibrium structure.
In Table \ref{tab:energiesN2}, the electronic energies determined in CASSCF, CASPT2, DMRG,
and CC calculations are summarized. For a better overview, the electronic energies at different internuclear distances are depicted in Figure \ref{fig:n2-energies} 
(all energies calculated with different methods at the equilibrium distance have been chosen as the zero-Hartree reference).

\begin{table}[h]
\caption{Electronic energies for the N$_2$ molecule in Hartree for CASSCF, CASPT2,
DMRG, CCSD, CCSD(T), and CCSDT calculations at different interatomic distances $d_{\rm NN}$.
}\label{tab:energiesN2}
{\footnotesize
\begin{center}
\begin{tabular}{p{1.9cm}cccccc}\hline \hline
Method   & $1.12$~\AA{}      &  $1.69$~\AA{}   & $2.12$~\AA{} &  $ 2.22$~\AA{}      & $2.33$~\AA{} & $ 3.18$~\AA{} \\\hline
CAS(10,8)SCF           &$-$109.132 549 &$-$108.882 463  &$-$108.804 469  &$-$108.800 108   &$-$108.797 850 &$-$108.795 397\\
CAS(10,8)PT2           &$-$109.359 217 &$-$109.123 008  &$-$109.034 004  &$-$109.025 152   &$-$109.019 396 &$-$109.008 111 \\
       &&& \\                
DMRG(10,46)            &$-$109.229 813 &$-$108.995 701	&$-$108.912 849	 &$-$108.905 589   &$-$108.900 737 &$-$108.881 603\\       
       &&&&& \\
CCSD(10,47)	&$-$109.000 679	& $-$108.941 650	&$-$108.909 788	&$-$108.928 743 &$-$108.950 482	&$^{\rm a}$ \\
CCSD(T)(10,47)&$-$109.000 713	&$-$108.985 858	&$-$109.041 195 	&$-$109.110 778	&$-$109.187 441	&$^{\rm a}$  \\
CCSDT(10,47) &$-$109.006 569	&$-$108.986 784	&$-$108.986 784	&$-$109.038 483	&$-$109.055 259	&$^{\rm a}$\\ 
      &&& \\          
%OLD (1s not frozen)
%CCSD(10,all)           &$-$109.389 768 & $-$109.093 635 &$-$108.972 810  &$-$108.972 720   &$-$108.983 272 &$^{\rm a}$  \\
%CCSD(T)(10,all)        &$-$109.410 265 & $-$109.151 979 &$-$109.113 345  &$-$109.162 841   &$-$109.236 466 &$^{\rm a}$  \\
%CCSDT(10,all)          &$-$109.409 872 & $-$109.148 686 &$-$109.144 884  &$-$109.167 927   &$^{\rm a}$  &$^{\rm a}$ \\
%
%NEW (1s frozen)
CCSD(10,all)    &$-$109.360 160 &$-$109.070 979	&$-$108.953 165	&$-$108.954 475	&$-$108.966 565&$^{\rm a}$  \\
CCSD(T)(10,all)	&$-$109.380 205	&$-$109.128 573	&$-$109.095 725	&$-$109.144 281	&$-$109.222 288&$^{\rm a}$  \\
CCSDT(10,all)	&$-$109.379 815	&$-$109.124 975 &$-$109.122 251	&$-$109.144 221	&$-$109.162 828&$^{\rm a}$   \\
\hline \hline
\end{tabular}
\begin{tablenotes}\footnotesize
\item $^{\rm a}$ Not computed due to convergence difficulties.\\
\end{tablenotes}
\end{center}
}
\end{table}

The restricted-virtual CC calculations yield, in general, higher electronic energies than
the DMRG(10,46) calculations for short and intermediate internuclear distances.
To describe correlation effects in the N$_2$ molecule appropriately,
higher order excitation operators in the cluster operator are mandatory \cite{Chan_N2}.
We should note that the dimension of the active space had to be enlarged by one
additional virtual orbital
in all restricted-virtual CC calculations to prevent symmetry breaking in the doubly-degenerate $\phi_g$-orbitals (b$_{2g}\oplus$ b$_{3g}$) and hence to avoid an unphysical lowering of the electronic
energy.

\begin{figure}[H]
\centering
\includegraphics[width=0.8\linewidth]{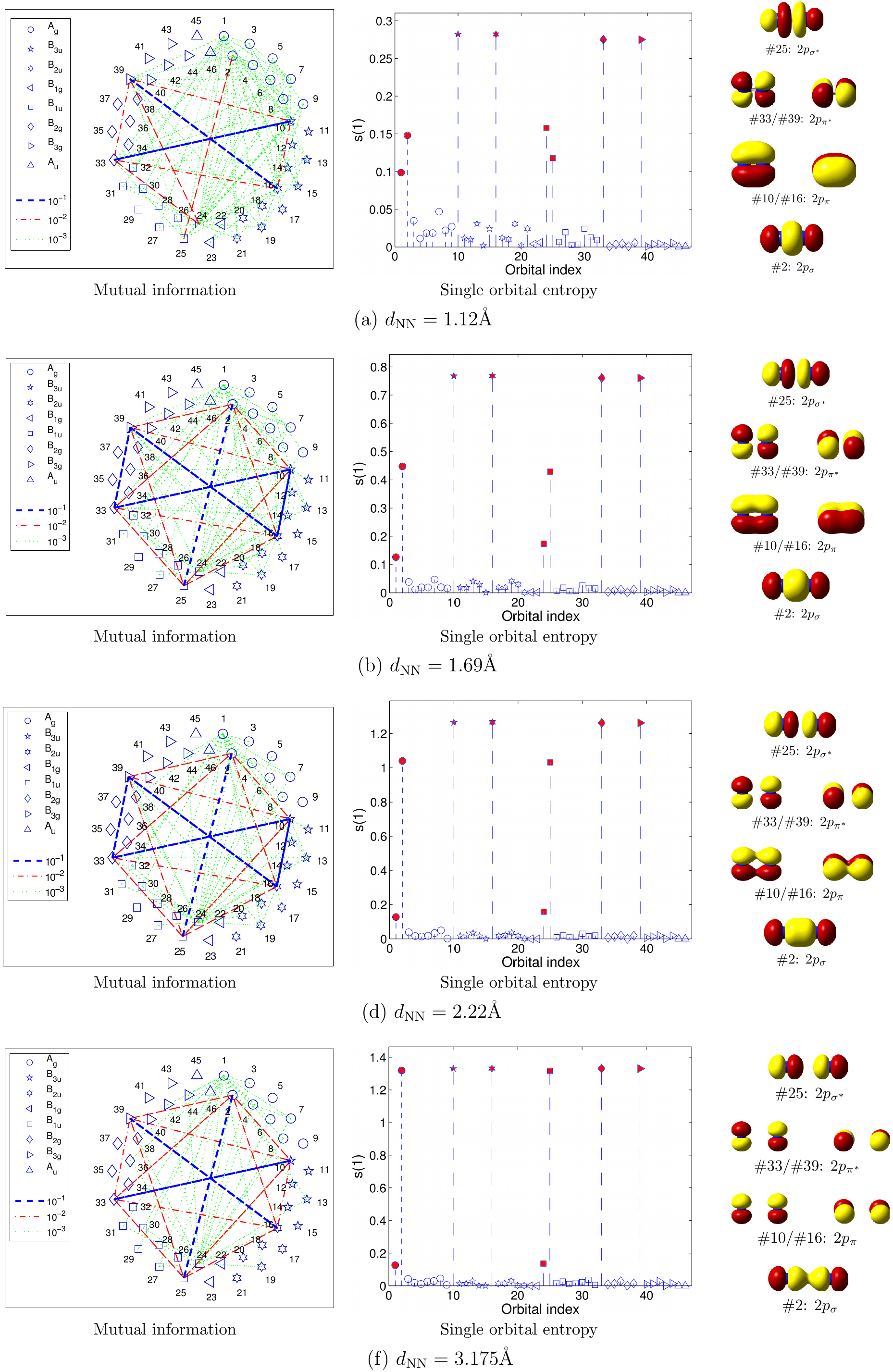}
\caption{\small
Mutual information and single-orbital entropies $s(1)_i$ from DMRG(10,46)[512,1024,$10^{-5}$]
calculations for the N$_2$ molecule at different internuclear distances.
The orbitals are numbered and sorted according to their (CASSCF) natural occupation numbers.
Strongly entangled orbitals are shown on the right hand side.
Each orbital index in the $s(1)_i$ diagram (middle; those included in the CAS(10,8)SCF calculations are marked in red) corresponds
to the same natural orbital as numbered in the entanglement plot (left).
The total quantum information $I_{\text{tot}}$ is 2.09, 4.80, 8.00, and 8.77 with increasing distance.
}\label{fig:n2}
\end{figure}

\begin{figure}[H]
\centering
\includegraphics[width=0.65\linewidth]{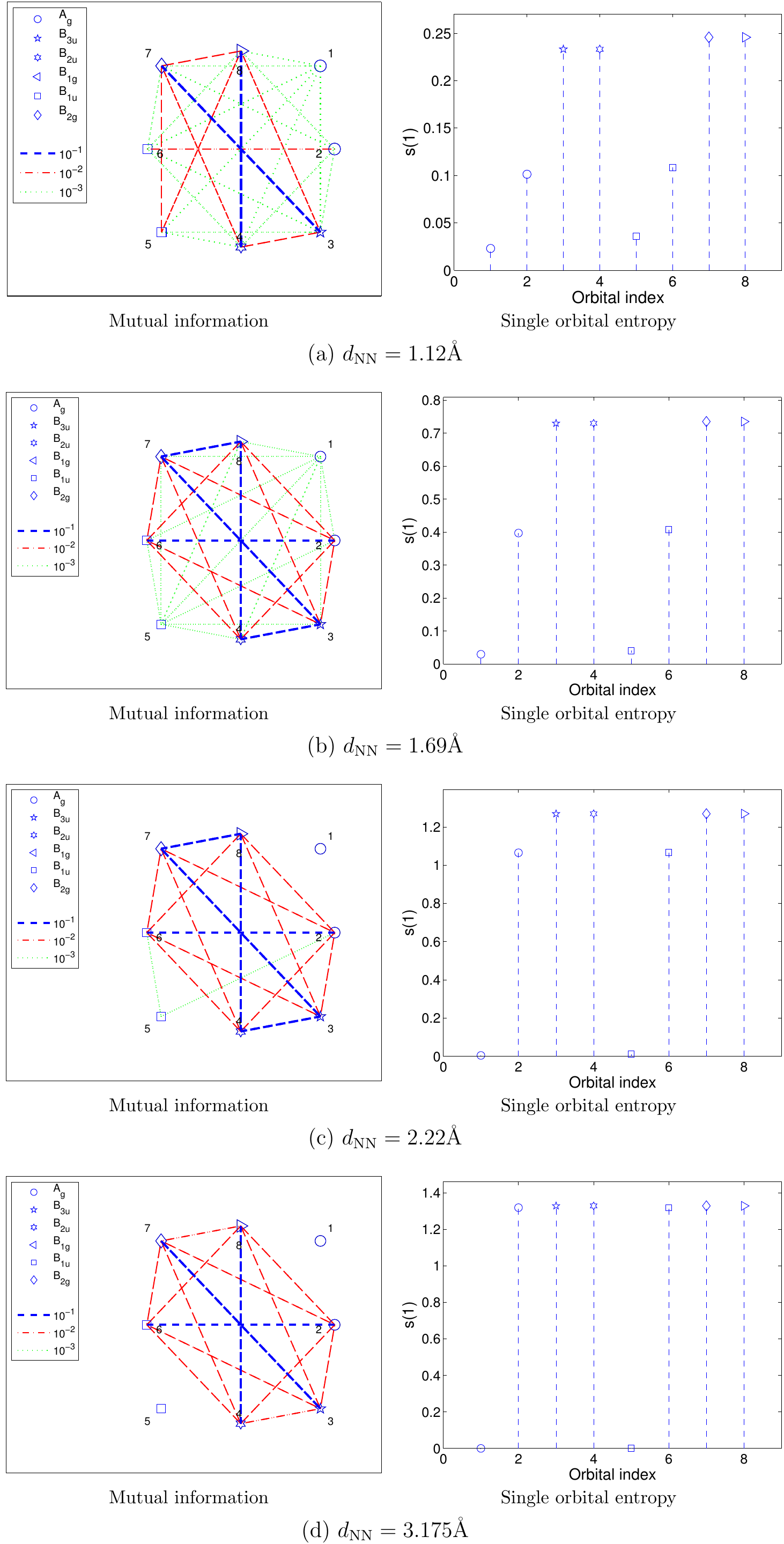}
\caption{\small
Mutual information (left) and single-orbital entropies $s(1)_i$ (right) from DMRG(10,8) calculations for the N$_2$ molecule at different internuclear distances.
The orbitals --- marked in red in Figure \ref{fig:n2} --- are numbered and sorted according to their (CASSCF) natural occupation numbers. 
The orbital index and the number in the entanglement plot (left) correspond to the same natural orbital.
}\label{fig:n2-cas}
\end{figure}

The slope of the potential energy surface shown in Figure \ref{fig:n2-energies} is either
too steep (CCSD(10,all)) or
too flat (CCSD(T) and CCSDT) and declines unphysically from an internuclear distance of 
2.10~\AA{} onwards. Restricting the number of active virtual orbitals even 
underestimates bonding
and amplifies the unphysical behavior in the electronic energy at longer bond lengths.
Similar observations were reported by other authors \cite{Piecuch_N2,Chan_N2}.
Note that it was not possible to converge CC calculations when
approaching the dissociation limit
(larger than and including 2.33 \AA{}), most probably due to the large atomic basis 
sets employed in this study \cite{Chan_N2}. 
It is worth mentioning that the CAS(10,8)SCF and CAS(10,8)PT2 results
agree well with the DMRG(10,46) reference calculations.

The different performance of CC, CASSCF, and CASPT2 calculations can be explained by
exploring the single-orbital entropy and mutual information diagrams shown in Figure \ref{fig:n2}
for selected interatomic distances (additional entanglement diagrams can be found in the
Supporting Information).
Close to the equilibrium structure, both $\pi$- and $\pi^*$-orbitals (\#10-\#33,\#16-\#39) are
strongly entangled, followed by the bonding and antibonding combinations of the $\sigma$-orbitals
(\#2-\#25), while all remaining orbitals are important to capture dynamic electron correlation
effects. When the nitrogen atoms are pulled apart, the single-orbital entropies corresponding
to the $\sigma, \sigma^*, \pi$, and $\pi^*$-orbitals increase considerably. However, we
still observe a large number of orbitals which are dynamically entangled. This explains
the qualitatively good performance of the CCSD(T)(14,all) and CCSDT(14,all) calculations
close to the equilibrium structure and for small internuclear distances
compared to the DMRG reference (see Figure \ref{fig:n2-energies}).
However, the amount of dynamic correlation decreases upon dissociation, and the
system becomes dominated by static and nondynamic electron correlation
(note the decreasing number of
green lines and increasing number of single-orbital entropies close to zero with increasing distances).
Thus, the standard single-reference CC fails---as expected---in describing the dissociation process of the N$_2$ molecule \cite{Piecuch_N2,Chan_N2}.

Figure \ref{fig:n2-cas} shows the entanglement diagrams for a DMRG(10,8) calculation,
which is equivalent to the CAS(10,8)SCF result. For all internuclear distances, the single-orbital
entropies corresponding to the statically entangled orbitals are underestimated
(compare Figures \ref{fig:n2} and \ref{fig:n2-cas}), which can be partially explained by the missing
dynamic correlation effects attributed to the active and virtual orbitals. Although the differences
in static correlation decrease when the atoms are pulled apart, CASSCF yields a qualitatively wrong
entanglement diagram, where it progressively dilutes dynamic correlation involved in both
nonbonding N $2s$-orbitals (\#1 and \#5 in Figure \ref{fig:n2-cas}),
\emph{i.e.}, the number of green lines and the single-orbital entropies diminish.
If the missing dynamic correlation effects are to be captured \emph{a posteriori}, for instance,
by means of perturbation theory, a larger amount of dynamic correlation needs to be
included close to the equilibrium structure and for smaller bond lengths than asymptotically, when approaching the
dissociation limit. This may explain why the CASPT2 dissociation curve deviates more strongly from
a FCI reference around the equilibrium bond length as presented in Ref.\ \onlinecite{CASPT2_N2}. 

Last but not least, we discuss how the entanglement diagrams can be utilized to monitor the
bond-forming/bond-breaking process and to resolve different types of bonds individually.
In the case of the N$_2$ molecule, we should be able to distinguish the dissociation of
two $\pi$-bonds and one $\sigma$-bond. If the N atoms are pulled apart,
the single-orbital entropies corresponding to the $\pi$- and $\pi^*$-orbital pairs
increase considerably faster than those corresponding to the bonding and antibonding 
combination of $\sigma$-orbitals. Thus, the weaker $\pi$-bonds are breaking
first under dissociation,
followed by the stronger $\sigma$-bond. If the nitrogen atoms are pulled further apart,
from a distance of approximately $1.6$ \AA{} onwards,
the $\sigma$-bond gets weakened and the corresponding single-orbital entropies increase
most extensively, while the $s(1)_i$ values associated with the $\pi$-bonds grow more
slowly. In the dissociation limit, where both the $\sigma$- and $\pi$-bonds are broken,
the single-orbital entropies corresponding to the bonding and antibonding combination
reach their maximum value of $\ln 4$ (note that $\langle\hat{n}_{\uparrow,i}\rangle = \langle\hat{n}_{\downarrow,i}\rangle= 0.5$ 
and $\langle\hat{n}_{\uparrow,i}\hat{n}_{\downarrow,i}\rangle=0.25$; see Ref.\ \onlinecite{Rissler2006519} for further discussion), 
first observed for the weaker $\pi$-bonds, followed by the stronger $\sigma$-bond.

\subsection{The fluorine molecule} 

The dissociation of the weakly covalently bonded F$_2$ molecule is a prime example of a
single-reference problem where dynamic electron correlation effects play a dominant role
\cite{Jankowski_F2,Ahlrichs_bs,Paldus_F2,Piecuch_F2,Ivanov_F2,Monika_F2}. 
In fact, a full valence CASSCF calculation yields only half of the binding energy \cite{Laidig_F2_N2},
while CCSD produces a potential energy well which is almost twice as deep as CCSDT
\cite{Piecuch_F2}.

\begin{figure}[H]
\centering
\includegraphics[width=0.8\linewidth]{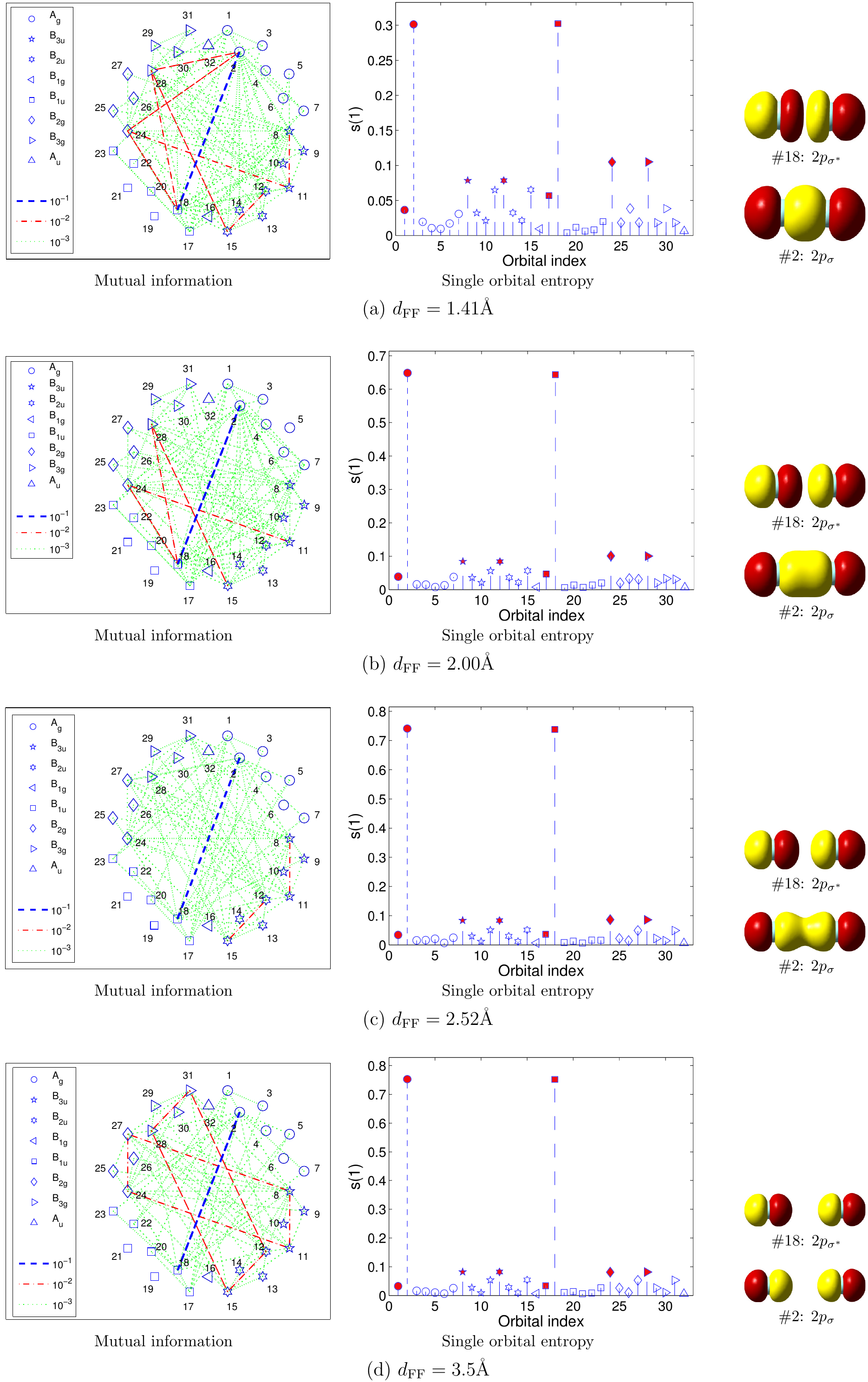}
\caption{\small
Mutual information and single-orbital entropies $s(1)_i$ from DMRG(14,32) calculations for the
F$_2$ molecule at different internuclear distances. 
The orbitals are numbered and sorted according to their (CASSCF) natural occupation numbers.
Strongly entangled orbitals are shown on the right hand side.
Each orbital index in the $s(1)_i$ diagram (middle; those included in the CAS(14,8)SCF calculations are marked in red) corresponds
to the same natural orbital as numbered in the entanglement plot (left).
%The total quantum information 
$I_{\text{tot}}$ is 1.41, 2.30, 2.40, and 2.41 with increasing distance.
}\label{fig:f2-dmrg}
\end{figure}

The importance of dynamic correlation effects can be clearly seen 
in the distribution of entanglement bonds (connecting lines in the mutual information plot) among orbitals
(see left column in Figure \ref{fig:f2-dmrg}). For the entanglement analysis,
we chose four characteristic points along the dissociation coordinate: the equilibrium
structure (1.41 \AA{}), two stretched bond length (2.00 and 2.53 \AA{}), and a bond length in the vicinity
of full dissociation (3.50 \AA{}).

\begin{figure}[H]
\centering
\includegraphics[width=0.9\linewidth]{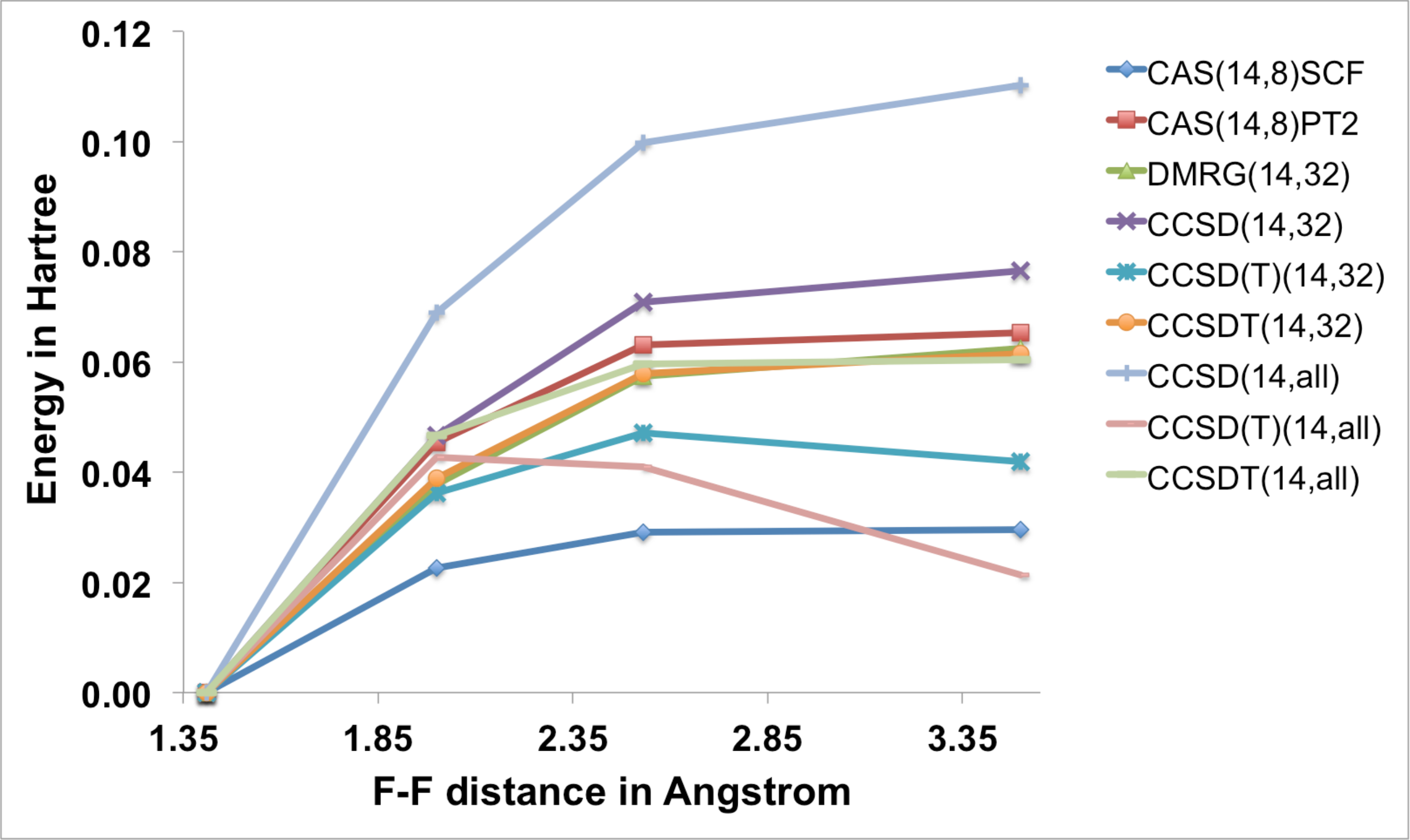}
\caption{
Electronic energy differences for the F$_2$ molecule at various intermolecular distances determined
by different quantum chemical methods. The energy reference is the electronic energy at equilibrium distance of each method.
}\label{fig:f2-energies}
\end{figure}

Table~\ref{tbl:CC_F2} collects the electronic energies for selected interatomic
distances determined by CC, CASSCF, CASPT2, and DMRG calculations. In general, the
DMRG(14,32) calculations yield rather similar (but, of course, slightly lower) electronic energies than
CCSDT(14,32).
%MR: Just out of curiosity: I am wondering how much of this lowering is actually due to the CAS orbitals
In particular, CAS(14,8)PT2, CCSDT(14,32), and DMRG(14,32) show a qualitatively similar asymptotic behavior (see Figure \ref{fig:f2-energies}).

\begin{table}[H]
\caption{Electronic energies for the F$_2$ molecule in Hartree for the CASSCF, 
CASPT2, DMRG, CCSD, CCSD(T) and CCSDT 
calculations at different interatomic distances $d_{\rm FF}$.
The DMRG parameter sets are summarized in the Supporting Information.}\label{tbl:CC_F2}
{
\begin{center}
\begin{tabular}{lccccc}\hline \hline
%         & \multicolumn{4}{c}{Energy in Hartree } \\ \cline{2-5}
Method   & $1.41$~\AA{}  & $2.00 $~\AA{}     &  $2.53$~\AA{}     & $3.50$~\AA{}    \\\hline
CAS(14,8)SCF    &$-$198.833 651&$-$198.810 987&$-$198.804 463  & $-$198.804 006\\
CAS(14,8)PT2    &$-$199.295 240&$-$199.249 765&$-$199.232 141  & $-$199.229 908   \\
       &&& \\[-3ex]     
DMRG(14,32)    &$-$198.968 889&$-$199.931 106  & $-$198.911 450 & $-$198.906 415  \\ 
%DMRG(14,58)       &$-$		& $-$199.027 5xx& $-$199.022 xxx	   \\                        
       &&& \\[-3ex]                                             
CCSD(14,32)    &$-$198.960 881&$-$198.914 147& $-$198.890 009& $-$198.884 283\\
CCSD(T)(14,32) &$-$198.967 091&$-$198.930 853& $-$198.919 979& $-$198.925 141\\
CCSDT(14,32)   &$-$198.967 501&$-$198.928 587&$-$198.909 625& $-$198.905 900\\ 
       &&& \\[-3ex] 
%OLD: 1s was not frozen
%CCSD(14,all)    &$-$199.320 352 &$-$199.224 737 & $-$199.217 581& $-$199.206 870   \\
%CCSD(T)(14,all) &$-$199.340 572 &$-$199.270 842 & $-$199.297 244 & $-$199.316 991   \\
%CCSDT(14,all)   &$-$199.340 580 &$-$199.266 962 & $-$199.253 970 & $-$199.253 149 \\
%
%NEW: 1s is frozen
CCSD(14,all)	&$-$199.293 721	&$-$199.224 737	&$-$199.193 891	&$-$199.183 541\\
CCSD(T)(14,all)	&$-$199.313 606	&$-$199.270 842	&$-$199.272 581	&$-$199.292 289\\
CCSDT(14,all) 	&$-$199.313 616	&$-$199.266 962	&$-$199.253 970	&$-$199.253 149 \\
\hline \hline
\end{tabular}
\end{center}
}
\end{table}

Figure \ref{fig:f2-dmrg} depicts the entanglement
measures determined from the DMRG(14,32) reference calculations. As expected, all orbitals are
involved in pure dynamic correlation (small single-orbital entropies and they are connected by
green lines in the mutual information diagram), except for the bonding and antibonding
$2p_{\sigma}$ combinations (\#2 and \#18) forming the single bond, which are more strongly entangled (large
single-orbital entropies and connected by a blue line encoding nondynamic correlation effects). Note that some orbitals included in a
full-valence CASSCF calculation (marked in red in the middle panel of Figure \ref{fig:f2-dmrg}) possess considerably
smaller $s(1)_i$ than some virtual orbitals, and are thus less important for dynamic correlation
effects.
If the atoms are pulled apart, the $\sigma$-bond starts to break, which is accompanied by an increase in the corresponding single-orbital entropies. For the remaining active
space orbitals, however, the single-orbital entropy patterns remain unchanged (note the different
scaling of the axes in the single-orbital entropy diagrams), while simultaneously
a larger number of orbitals becomes weakly entangled (increasing number of green lines
in Figure \ref{fig:f2-dmrg}).

This can also be observed in the evolution of the total quantum
information and its contributions.
Close to the equilibrium distance, weakly entangled orbitals account for the
essential part of the total quantum information, while statically/nondynamically
entangled orbitals contribute most
for stretched interatomic distances. Furthermore, the total quantum information
accumulates upon dissociation, which motivates the increase in both static (predominantly)
and dynamic (secondary) correlation effects. 
Our entanglement-based analysis thus indicates that a restricted CC calculation should
describe the dissociation process of the F$_2$ molecule properly, since all active space 
orbitals are mainly dynamically entangled, despite one strongly entangled pair of orbitals.

\begin{figure}[H]
\centering
\includegraphics[width=0.6\linewidth]{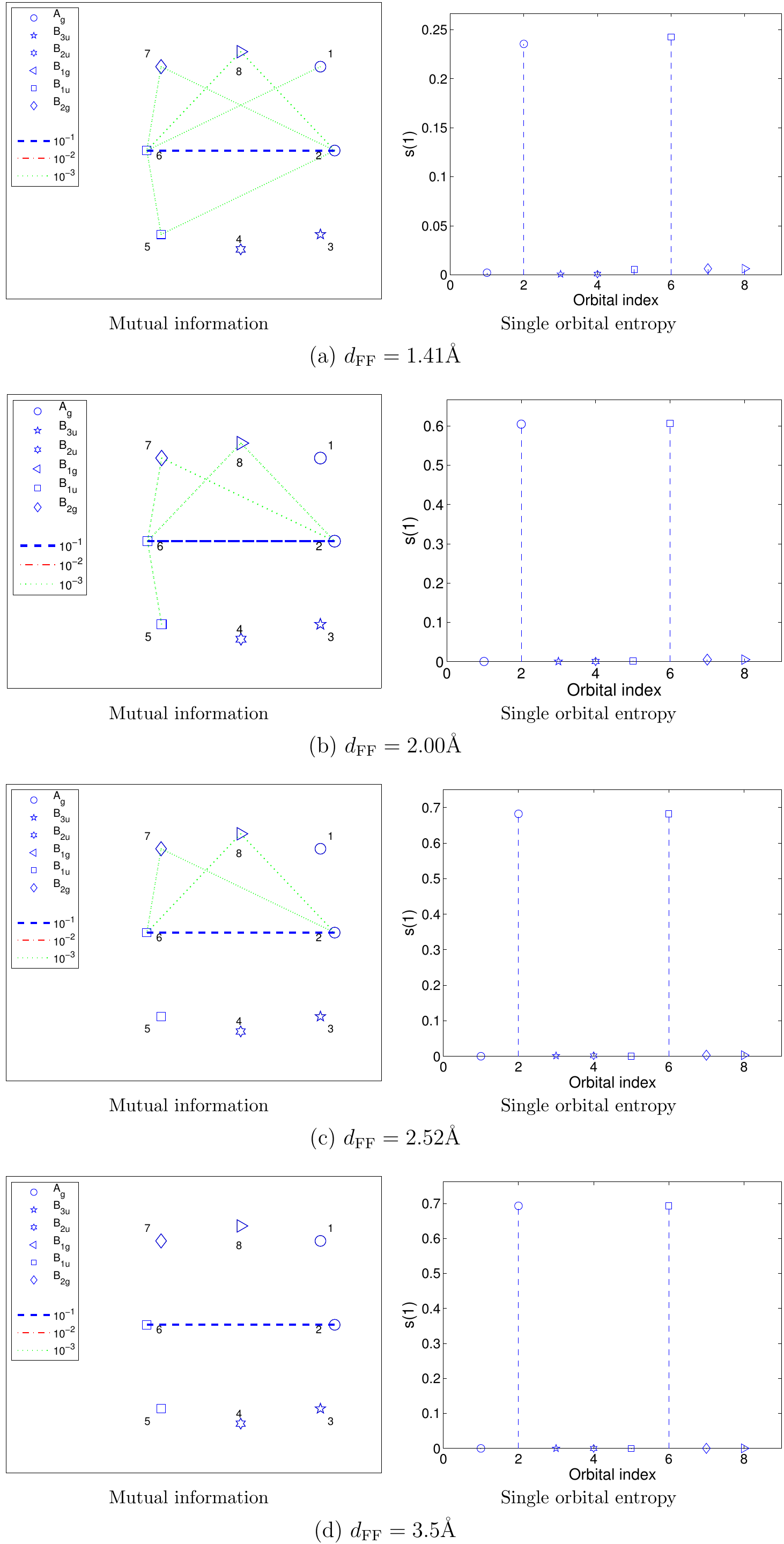}
\caption{\small
Mutual information (left) and single-orbital entropies $s(1)_i$ (right) from DMRG(14,8) calculations for the F$_2$ molecule at different internuclear distances.
The orbitals --- marked in red in Figure \ref{fig:f2-dmrg} --- are numbered and sorted according to their (CASSCF) natural occupation numbers. 
The orbital index and the number in the entanglement plot (left) correspond to the same natural orbital.
}\label{fig:f2-cas}
\end{figure}

%\sout{Note that it has been argued that the unrestricted CC formulation is sufficient to capture the static electron correlation
%effects attributed to the bonding and antibonding combination of the $2p_{\sigma}$-orbitals \cite{Monika_F2}.}
%
Figure \ref{fig:f2-cas} shows the entanglement diagram determined from a DMRG(14,8) calculation, which is equivalent to the CAS(14,8)SCF calculation. 
By comparing Figures \ref{fig:f2-dmrg} and \ref{fig:f2-cas}, we observe that a large amount of dynamic electron correlation effects cannot be 
captured in the small active space calculations (note the small number of green lines and single-orbital entropies close to zero in Figure \ref{fig:f2-cas}).
If the F atoms are pulled apart, the single-orbital entropies and mutual information will continue to decrease, which
implies that even less dynamic electron correlation could be captured in a CAS(14,8)SCF calculation
for increasing interatomic distances than for those close to the equilibrium bond length. In the dissociation limit, only the bonding and antibonding
$\sigma$-orbitals are strongly entangled, and a CAS(14,8)SCF wave function misses all
dynamic correlation effects among the active-space orbitals (no green lines and single-orbital entropies close to zero). 
This has direct consequences
for the calculation of the dissociation pathway of F$_2$. Since CAS(14,8)SCF neglects the
major part of dynamic electron correlation effects, it suffers from a wrong dissociation limit,
which is in accord with previous findings \cite{Laidig_F2_N2}.
Furthermore, since dynamic correlation effects are captured in an unbalanced way already in the 
CAS(14,8)SCF calculation (recall the decreasing number of green lines or vanishing values
of single-orbital entropies), a CAS(14,8)PT2 treatment cannot capture the missing dynamic
correlation in the active space, an observation that we have already at the example of N$_2$
dissociation. Hence, the CAS(14,8)PT2 dissociation curve comprises a different slope than the DMRG(14,32) or CCSDT(14,all) reference calculations in
Figure \ref{fig:f2-energies}.

Finally, we can monitor the bond-breaking process of a $\sigma$-bond in the F$_2$
molecule employing the entanglement analysis. In the equilibrium structure, the bonding
and antibonding combination of the F $2p_z$-orbitals feature medium-sized single-orbital
entropies and are thus statically entangled. If both atoms are pulled apart, only the
single-orbital entropies corresponding to the bonding and antibonding orbitals of
the $\sigma$-bond increase significantly from about 0.3 to 0.75.
If the F atoms are further pulled apart, the $s(1)_i$ profile will change only little indicating that
the $\sigma$-bond is almost broken. This is also seen in the flat slope of the
energy between 2.52 and 3.70 \AA{} in Figure \ref{fig:f2-energies}.

We should note that the maximum value of $s(1)_i$ of $\ln 4$ cannot be reached during
the bond-breaking (or bond-forming) process of one single-bond. This can be explained by the
structure of the electronic wave function and by how the $s(1)_i$ measure has been defined
(cf., e.g., Table \ref{tab:fermion-ops-1}). In the case of one single-bond, the matrix elements
of ${\cal O}_i^{(1)}$ and ${\cal O}_i^{(6)}$ are significantly smaller than those determined from 
electronic wave functions describing the dissociation of multiple or several single-bonds at once.
The qualitative picture, however,
remains unchanged.

\begin{figure}[H]
\centering
\includegraphics[width=0.9\linewidth]{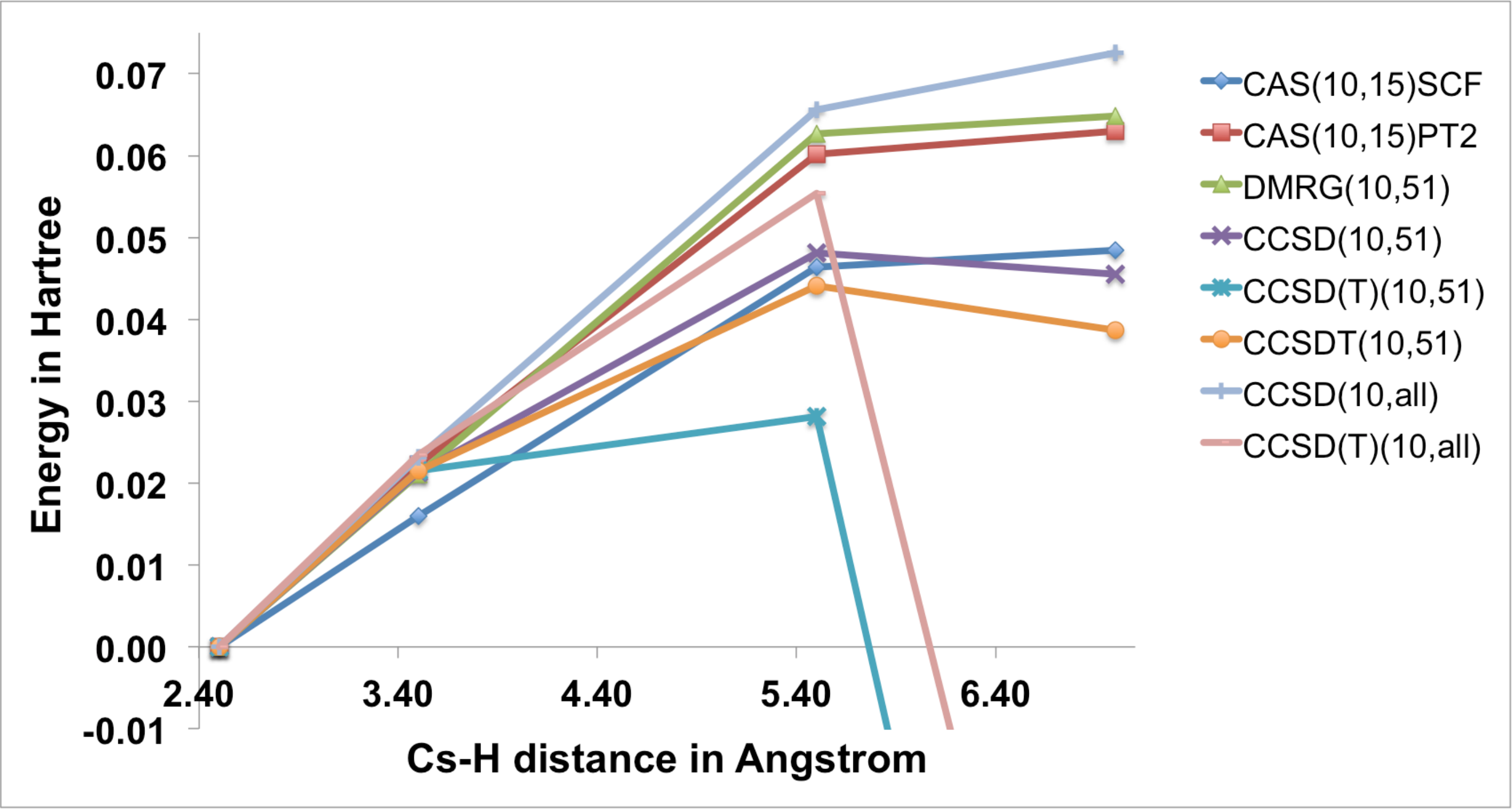}
\caption{
Electronic energy differences for the CsH molecule at various intermolecular distances determined
by different quantum chemical methods. The energy reference is the electronic energy at equilibrium distance of each method.
}\label{fig:csh-energies}
\end{figure}

\subsection{The cesium hydride molecule} 
The dissociation of CsH represents another prototypical quantum chemical problem, namely that of an avoided crossing \cite{Carnell_CsH,Reiher_CsH,Zrafi_2006a,Zrafi_2006b,Yan_CsH}.
Table \ref{tab:energiesCsH} summarizes the electronic energies determined from CASSCF, CASPT2,
DMRG and CC calculations for different active spaces. CAS(10,15)PT2 and DMRG(10,51)
yield qualitatively similar dissociation curves (they almost lie on top of each other in
Figure \ref{fig:csh-energies}), while all restricted-virtual CC electronic energies are significantly
above the DMRG and
CASPT2 results (\textit{cf.} Table \ref{tab:energiesCsH}) for short and intermediate internuclear distances.

\begin{threeparttable}[H]
\caption{Electronic energies for the CsH molecule in Hartree for CASSCF, CASPT2,
DMRG, CCSD$^{\rm a}$, CCSD(T)$^{\rm a}$ and CCSDT$^{\rm a}$ calculations at different interatomic distances $d_{\rm CsH}$.}\label{tab:energiesCsH}
\begin{center}
\begin{tabular}{lccccc}\hline \hline
Method          	& $2.50$~\AA{}   	&$3.50$~\AA{}     	&  $5.50$~\AA{}   	& $7.00$~\AA{} \\\hline
CAS(10,15)SCF	&$-$7783.942 879 &$-$7783.926 862 &$-$7783.896 474 &$-$7783.894 382  \\
CAS(10,15)PT2	&$-$7784.023 052 &$-$7784.000 788 &$-$7783.962 897 &$-$7783.960 022  \\
&&&& \\[-3ex]
 DMRG(10,51)	&$-$7783.971 143 &$-$7783.950 064	&$-$7783.908 452 &$-$7783.906 309   \\ 
&&&& \\[-3ex]
CCSD(10,51)	  &$-$7783.933 885&$-$7783.912603	&$-$	7783.885 783&$-$7783.888 379 \\
CCSD(T)(10,51)&$-$7783.937 106&$-$7783.915632	&$-$7783.908 993	&$-$7784.069 822 \\
CCSDT(10,51)	  &$-$7783.950 385&$-$7783.915985	&$-$	7783.893 401&$-$7783.898 877 \\
&&&& \\[-3ex]
CCSD(10,all)		&$-$7784.032498	&$-$7784.009 397	&$-$7783.966 855	&$-$7783.959 935 \\
CCSD(T)(10,all)	&$-$7784.039581	&$-$7784.016 104	&$-$7783.984 226	&$-$7784.130 504 \\
\hline \hline
\end{tabular}
\begin{tablenotes}\footnotesize
\item[a] In all CC calculations, the DKH3 HF reference energy has been shifted to
coincide with the DKH10 HF reference energy. The correlation energy remains unaffected by
the choice of the DKH Hamiltonian.\\
\end{tablenotes}
\end{center}
\end{threeparttable}

\begin{figure}[H]
\centering
\includegraphics[width=0.7\linewidth]{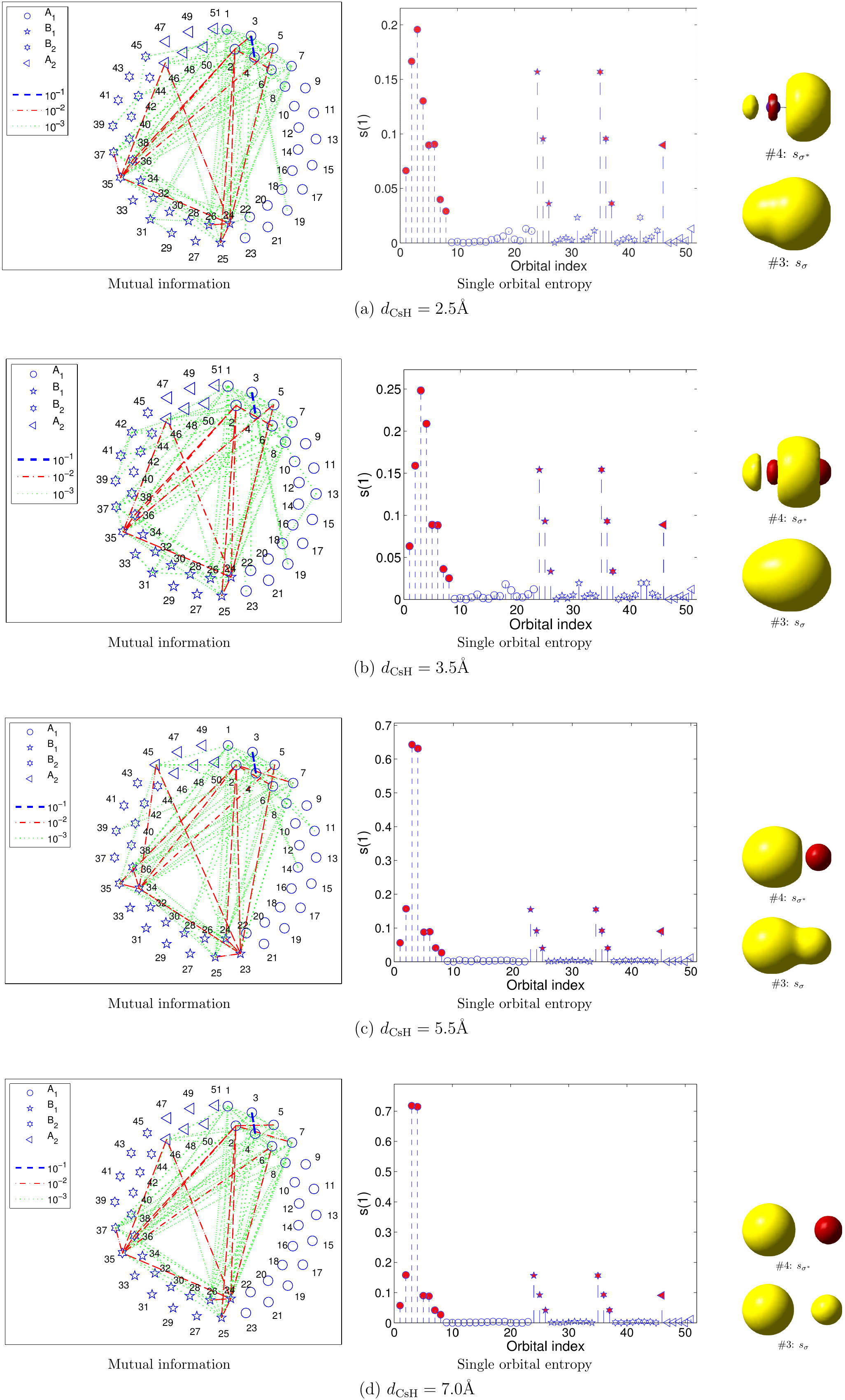}
\caption{\small
Mutual information and single-orbital entropies $s(1)_i$ from DMRG(10,51)[512,1024,$10^{-5}$]
calculations for the CsH molecule at different internuclear distances.
The orbitals are numbered and sorted according to their (CASSCF) natural occupation numbers.
Strongly entangled orbitals are shown on the right hand side (from left to right: Cs, H).
Each orbital index in the $s(1)_i$ diagram (middle; those included in the CAS(14,8)SCF calculations are marked in red) corresponds
to the same natural orbital as numbered in the entanglement plot (left).
The total quantum information $I_{\text{tot}}$ is 1.66, 1.75, 2.49, and 2.64 for increasing distance.
}\label{fig:csh}
\end{figure}

The entanglement diagrams along the dissociation pathway shown in Figure \ref{fig:csh} illustrate that the major part of the
static and dynamic correlation is distributed among the active-space orbitals already incorporated
in the CASSCF calculations. Only some few external orbitals contribute to dynamic correlation
effects. Upon dissociation, the bonding and antibonding $\sigma$-orbitals (\#3-\#4)
become strongly entangled, while the single-orbital entropies corresponding to the
external orbitals of the CAS(10,15) active space decrease, signifying a reduced amount
of dynamic correlation effects attributed to these external orbitals. This picture remains
unchanged if the atoms are continuously pulled apart, despite the further increase
in single-orbital entropies corresponding to the $\sigma$- and $\sigma^*$-orbitals.
As expected, those orbitals which are involved in chemical bonding become strongly
entangled when we reach the dissociation limit.

\begin{figure}[H]
\centering
\includegraphics[width=0.6\linewidth]{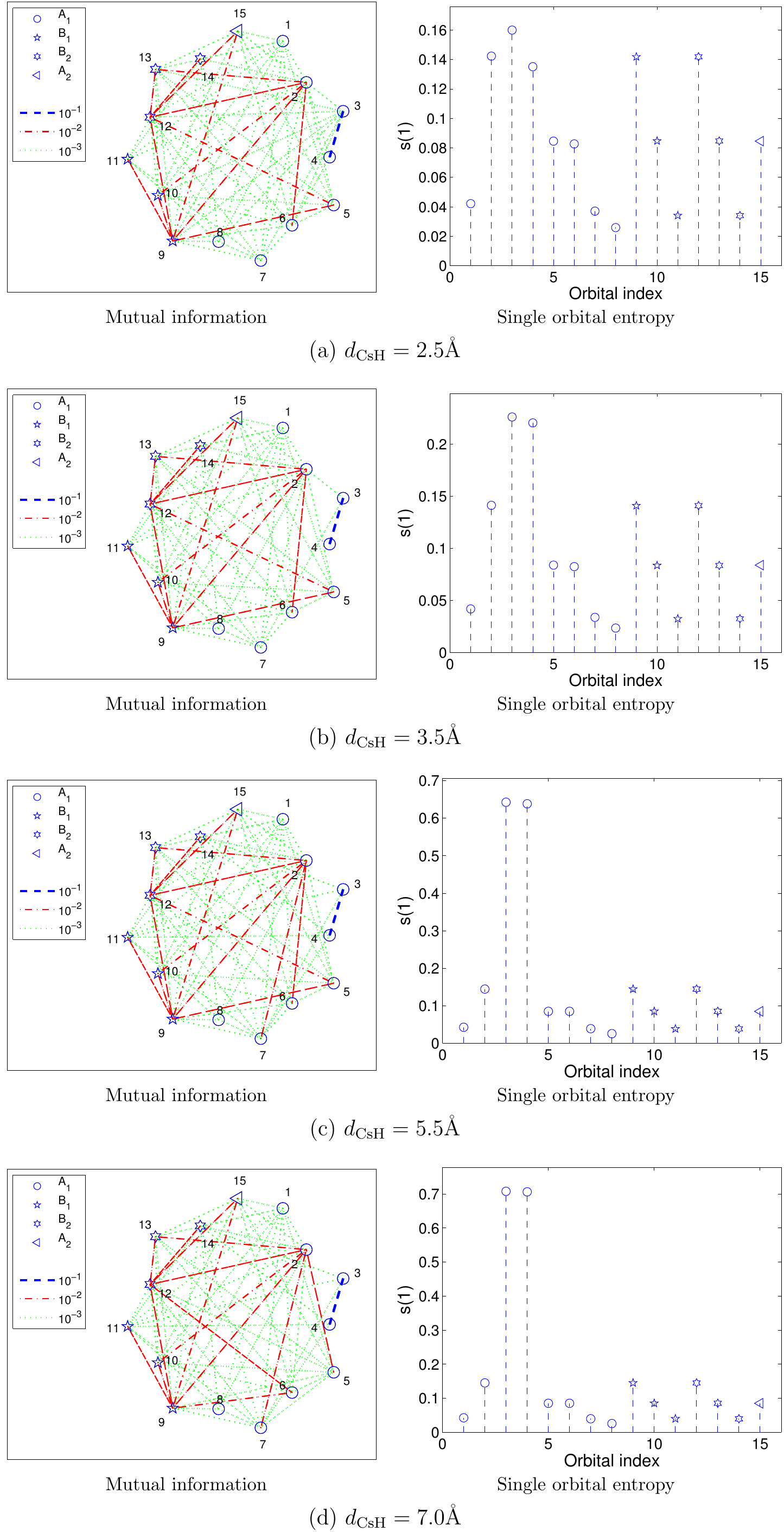}
\caption{\small
Mutual information and single-orbital entropies $s(1)_i$ from DMRG(10,15) calculations for the CsH
molecule at different internuclear distances.
The orbitals --- marked in red in Figure \ref{fig:csh} --- are numbered and sorted according to their (CASSCF) natural occupation numbers.
The orbital index and the number in the entanglement plot (left) correspond to the same natural orbital.
}\label{fig:csh-cas}
\end{figure}

If we compare the entanglement diagrams of the large active space calculation (Figure \ref{fig:csh})
to those determined for the small active space of CAS(10,15)SCF (Figure \ref{fig:csh-cas}), we observe similar entanglement
patterns for intermediate and stretched bond lengths. Minor differences in single-orbital
entropies are present around the equilibrium structure, where both static and dynamic correlation
effects are underestimated as indicated by $s(1)_i$ values being too small. Note that a similar
picture is expected for all remaining points of the potential energy curve.
Furthermore, since the major
part of electron correlation can be already appropriately included in a CAS(10,15)SCF calculation,
the missing dynamic electron correlation attributed to the (small number of) external
orbitals can be easily captured in a CASPT2 calculation. This explains the similarity in the
(slope of the) dissociation curves determined for CASPT2 and the DMRG calculation.
Note that DMRG yields overall higher electronic energies since the dynamic correlation effects
of the external orbitals are neglected, which results basically in a constant shift in energy
when compared to the CASPT2 results, which take the full virtual orbital space into account.

A comparison of the entanglement diagrams in Figures \ref{fig:csh} and \ref{fig:csh-cas} determined for small and large active space
calculations suggests that CAS(10,15)SCF is not able to incorporate all static and dynamic
correlation effects in the chosen active space in an equal manner along the whole reaction
coordinate. 
However, these differences are only minor and more pronounced around the equilibrium distance than for larger interatomic distances. Hence, the CAS(10,15)SCF
calculations lead to a qualitatively wrong dissociation pathway (recall the  steeper
slope in Figure \ref{fig:csh-energies}). 
Note that this can be easily corrected by employing perturbation theory upon the CAS(10,15)SCF wave function. 
Furthermore, since we observe a large number of statically entangled orbitals 
CCSD(T) yields a qualitatively and quantitatively incorrect
dissociation curve (see also Figure \ref{fig:csh-energies}).

As observed for the dissociation of the F$_2$ molecule, the single-orbital entropies associated
with the bonding and antibonding $\sigma$-orbitals forming the $\sigma$-bond increase
if the atoms are pulled apart, while those corresponding to the remaining orbitals change
only marginally. These small-valued $s(1)_i$ are caused by intra-atomic correlation effects
between the nonbonding Cs 4$d$-, 6$p$- ,5$d$- and 7$s$-orbitals.
The increase in $s(1)_i$ corresponding to the orbitals involved in bond-breaking during the
dissociation pathway leads to a gradual rise in the total quantum information, which would
reach its maximum value in the dissociation limit. The same could be observed for the
mutual information and single-orbital entropies.

\section{Conclusions}
In this work, we have elaborated on how our entanglement-based analysis
of electron-correlation effects introduced in
Ref.~\citenum{entanglement_letter} can be extended to study such effects in chemical reactions, i.e., in bond-making and bond-breaking processes. 
The mutual information and single-orbital
entropy are convenient
measures to resolve dynamic, static and nondynamic correlation effects
among molecular orbitals and to understand
the performance of \emph{ab initio} quantum chemical approaches.
The calculation of the one- and two-orbital reduced density matrices whose
eigenvalues enter the single-orbital entropy and mutual information, has been discussed in some detail.

We demonstrated that the one- and two-orbital entanglement measures can
be utilized to monitor bond-breaking and---equivalently---bond-forming processes.
Upon dissociation of a chemical bond, the bonding and antibonding molecular orbitals
associated with the bond of interest become strongly entangled. Hence, the corresponding
single-orbital entropies gradually increase if two atoms are pulled apart. Moreover,
the entanglement analysis resolves the bond breaking of different bond types ($\sigma$, $\pi$, etc.)
individually in multi-bonded centers. 
Whether it is possible to define a
quantitative bond strength employing entanglement measures requires, however, further
investigations.

As molecules with prototypical bonds, we have investigated the dissociation process of the 
diatomic molecules N$_2$, F$_2$ and CsH, which represent characteristic examples of single- or multi-reference
problems. 
For the N$_2$ molecule, our analysis indicates that nondynamic, static and dynamic correlation
effects are confined to a small number of orbitals which entails the qualitatively good
performance of
CASPT2-type approaches.
A well-known counter example represents the dissociation
problem of the F$_2$ molecule, where the entanglement measures illustrate that all active-space orbitals
contribute equally to dynamic correlation effects along the
reaction coordinate, while static correlation effects are confined to the bonding and antibonding
$\sigma$-orbitals only. In both cases, however, full-valence active space calculations considerably
underestimated dynamic correlation effects for stretched interatomic distances. 

A different picture was obtained for the dissociation of CsH, where
all important static and dynamic correlation effects could already be incorporated in
the small active space calculation, \emph{i.e.}, similar entanglement diagrams were obtained
in CASSCF and DMRG calculations. The missing dynamic correlation effects attributed to
the external CASSCF natural orbitals are minor and can thus be easily captured by
perturbation theory on top of the CASSCF reference function.

All bond-formation/bond-breaking processes discussed in this work have been studied for diatomic molecules only,
where the molecular orbitals are basically localized orbitals and point group symmetry could be exploited.
The transferability of the entanglement-based analysis to larger reactive systems, the effect of the type of
molecular orbitals chosen for the analysis, and the possible necessity for localizing  
molecular orbitals is currently under investigation in our laboratory. In this context,
the one- and two-orbital entanglement measures, calculated here from converged DMRG wave functions, 
should be implemented also for standard quantum chemical approaches (MP2, CC, CASSCF, etc.), which we will
reserve for future work.

\section*{Acknowledgments}
The authors gratefully acknowledge financial support by the Swiss national science foundation SNF (project 200020-144458/1),
and from the Hungarian Research Fund (OTKA) under Grant No.~K73455 and K100908.
{\"O}.L.\ acknowledges support from the Alexander von Humboldt foundation and from ETH Z\"urich
during his time as a visiting professor.

%\bibliographystyle{aip-mod}
%\bibliography{begin,literatura,end}

\begin{thebibliography}{10}

\bibitem{Lowdin_corr_energy}
P.-O. L\"owdin,
\newblock Phys. Rev. {\bf 97}, 1509 (1955).

\bibitem{Lowdin_rev}
P.-O. L\"owdin,
\newblock {Correlation Problem in Many-Electron Quantum Mechanics},
\newblock in {\em Adv. Chem. Phys.}, volume~I, chapter Review of Different
  Approaches and Discussion of Some Current Ideas, pages 209--321, Wiley \&
  Sons, Inc, 1958.

\bibitem{Bartlett_1994}
R.~J. Bartlett and J.~F. Stanton,
\newblock Rev. Comput. Chem. {\bf 5}, 65 (1994).

\bibitem{bartlett_2007}
R.~J. Bartlett and M.~Musia{\l},
\newblock Rev.~Mod.~Phys. {\bf 79}, 291 (2007).

\bibitem{Sinanoglu1963}
O.~Sinano\u{g}lu and D.~F. Tuan,
\newblock J. Chem. Phys. {\bf 38}, 1740 (1963).

\bibitem{Roos_casscf}
B.~Roos and P.~R. Taylor,
\newblock Chem. Phys. {\bf 48}, 157 (1980).

\bibitem{caspt21}
K.~Andersson, P.-A. Malmqvist, B.~O. Roos, A.~J. Sadlej, and K.~Woli{\'n}ski,
\newblock J. Chem. Phys. {\bf 94}, 5483 (1990).

\bibitem{caspt22}
K.~Andersson, P.-A. Malmqvist, and B.~O. Roos,
\newblock J. Chem. Phys. {\bf 96}, 1218 (1992).

\bibitem{Pulay_CASPT2_2011}
P.~Pulay,
\newblock Int. J. Quantum Chem. {\bf 111}, 3273 (2011).

\bibitem{adamowicz-mrcc}
V.~V. Ivanov, D.~I. Lyakh, and L.~Adamowicz,
\newblock Phys. Chem. Chem. Phys. {\bf 11}, 2355 (2009).

\bibitem{Bogus_MRCC}
B.~Jeziorski,
\newblock Mol. Phys. {\bf 108}, 3043 (2010).

\bibitem{monika_mrcc}
D.~I. Lyakh, M.~Musia{\l}, V.~F. Lotrich, and J.~Bartlett,
\newblock Chem. Rev. {\bf 112}, 182 (2012).

\bibitem{Koehn2013}
A.~K\"ohn, M.~Hanauer, L.~A. M\"uck, T.-C. Jagau, and J.~Gauss,
\newblock WIREs Comp. Mol. Sci. {\bf 3}, 176 (2013).

\bibitem{MR_test_Wilson}
W.~Jiang, N.~J. DeYonker, and A.~K. Wilson,
\newblock J. Chem. Theory Comput. {\bf 8}, 460 (2012).

\bibitem{lee_1989}
T.~J. Lee and P.~R. Taylor,
\newblock Int.~J.~Quantum~Chem. {\bf 23}, 199 (1989).

\bibitem{Lee_t1_1}
T.~J. Lee, J.~E. Rice, G.~E. Scuseria, and H.~F. {Schaefer III},
\newblock Theor. Chim. Acta {\bf 75}, 81 (1989).

\bibitem{T1_open-shell}
M.~L. Leininger, I.~M.~B. Nielsen, T.~D. Crawford, and C.~L. Janssen,
\newblock Chem. Phys. Lett. {\bf 328}, 431 (2000).

\bibitem{Lee_T1_PT}
T.~J. Lee, M.~Head-Gordon, and A.~P. Rendell,
\newblock Chem. Phys. Lett. {\bf 243}, 402 (1995).

\bibitem{Nielsen1999}
I.~M.~B. Nielsen and C.~L. Janssen,
\newblock Chem. Phys. Lett. {\bf 310}, 568 (1999).

\bibitem{Lee_T1_and_D1}
T.~J. Lee,
\newblock Chem. Phys. Lett. {\bf 372}, 362 (2003).

\bibitem{Ziesche_1995}
P.~Ziesche,
\newblock Int. J. Quantum Chem. {\bf 56}, 363 (1995).

\bibitem{Luzanov_2005}
A.~V. Luzanov and O.~V. Prezhdo,
\newblock Int. J. Quantum Chem. {\bf 102}, 582 (2005).

\bibitem{Huang2005}
Z.~Huang and S.~Kais,
\newblock Chem. Phys. Lett. {\bf 413}, 1 (2005).

\bibitem{Huang2006}
Z.~Huang, H.~Wang, and S.~Kais,
\newblock J. Mod. Opt. {\bf 53}, 2543 (2006).

\bibitem{Juhasz_2006}
T.~Juh\'{a}sz and D.~A. Mazziotti,
\newblock J. Chem. Phys. {\bf 125}, 174105 (2006).

\bibitem{Luzanov_2007}
A.~V. Luzanov and O.~Prezhdo,
\newblock Mol. Phys. {\bf 105}, 2879 (2007).

\bibitem{Kais2007}
S.~Kais,
\newblock {\em Entanglement, Electron Correlation, and Density Matrices}, pages
  493--535,
\newblock John Wiley \& Sons, Inc., 2007.

\bibitem{Greenman_2010}
L.~Greenman and D.~A. Mazziotti,
\newblock J. Chem. Phys. {\bf 133}, 164110 (2010).

\bibitem{Alcoba_2010}
D.~R. Alcoba, R.~C. Bochicchio, L.~Lain, and A.~Torre,
\newblock J. Chem. Phys. {\bf 133}, 144104 (2010).

\bibitem{Pelzer_2011}
K.~Pelzer, L.~Greenman, G.~Gidofalvi, and D.~A. Mazziotti,
\newblock J. Phys. Chem. A {\bf 115}, 5632 (2011).

\bibitem{Ivanov_2005}
V.~V. Ivanov, D.~I. Lyakh, and L.~Adamowicz,
\newblock Mol. Phys. {\bf 103}, 2131 (2005).

\bibitem{Takatsuka_1978}
K.~Takatsuka, T.~Fueno, and K.~Yamaguchi,
\newblock Theor. Chim. Acta {\bf 183}, 175 (1978).

\bibitem{Staroverov_2000}
V.~N. Staroverov and E.~R. Davidson,
\newblock Chem. Phys. Lett. {\bf 330}, 161 (2000).

\bibitem{Bochicchio_2003}
R.~Bochicchio, A.~Torre, and L.~Lain,
\newblock Chem. Phys. Lett. {\bf 380}, 486 (2003).

\bibitem{Head-Gordon_2003}
M.~Head-Gordon,
\newblock Chem. Phys. Lett. {\bf 380}, 488 (2003).

\bibitem{Hachmann_2007}
J.~Hachmann, J.~J. Dorando, M.~Avil\'{e}s, and G.~K.-L. Chan,
\newblock J. Chem. Phys. {\bf 127}, 134309 (2007).

\bibitem{entanglement_letter}
K.~Boguslawski, P.~Tecmer, O.~Legeza, and M.~Reiher,
\newblock J. Phys. Chem. Lett. {\bf 3}, 3129 (2012).

\bibitem{legeza_dbss}
O.~Legeza and J.~S\'olyom,
\newblock Phys. Rev. B {\bf 68}, 195116 (2003).

\bibitem{Legeza2006}
O.~Legeza and J.~S\'olyom,
\newblock Phys. Rev. Lett. {\bf 96}, 116401 (2006).

\bibitem{Rissler2006519}
J.~Rissler, R.~M. Noack, and S.~R. White,
\newblock Chem. Phys. {\bf 323}, 519 (2006).

\bibitem{legeza_dbss3}
O.~Legeza and J.~S\'olyom,
\newblock Phys. Rev. B {\bf 70}, 205118 (2004).

\bibitem{Barcza2013}
G.~Barcza, R.~M. Noack, J.~S\'o{}lyom, and O.~Legeza,
\newblock Entanglement topology of strongly correlated systems,
  Korrelationstage MPIPKS, Dresden  (2011).

\bibitem{scholl05}
U.~Schollw\"ock,
\newblock Rev. Mod. Phys. {\bf 77}, 259 (2005).

\bibitem{ors_springer}
O.~Legeza, R.~Noack, J.~S\'olyom, and L.~Tincani,
\newblock {Applications of Quantum Information in the Density-Matrix
  Renormalization Group},
\newblock in {\em Computational Many-Particle Physics}, edited by H.~Fehske,
  R.~Schneider, and A.~Wei\ss{}e, volume 739 of {\em Lect. Notes Phys.}, pages
  653--664, Springer, Berlin/Heidelerg, 2008.

\bibitem{marti2010b}
K.~H. Marti and M.~Reiher,
\newblock Z. Phys. Chem. {\bf 224}, 583 (2010).

\bibitem{chanreview}
G.~K.-L. Chan and S.~Sharma,
\newblock Annu. Rev. Phys. Chem. {\bf 62}, 465 (2011).

\bibitem{white}
S.~R. White,
\newblock Phys. Rev. Lett. {\bf 69}, 2863 (1992).

\bibitem{ANO-RCC_Cs}
V.~Veryazov and P.-O. Widmark,
\newblock Theor.~Chem.~Acc. {\bf 111}, 345 (2003).

\bibitem{DKH2}
B.~A. Hess,
\newblock Phys.~Rev.~A {\bf 33}, 3742 (1986).

\bibitem{Wolf_2002}
A.~Wolf, M.~Reiher, and B.~A. Hess,
\newblock J. Chem. Phys. {\bf 117}, 9215 (2002).

\bibitem{Reiher_2004a}
M.~Reiher and A.~Wolf,
\newblock J. Chem. Phys. {\bf 121}, 2037 (2004).

\bibitem{Reiher_2004b}
M.~Reiher and A.~Wolf,
\newblock J. Chem. Phys. {\bf 121}, 10945 (2004).

\bibitem{Siegbahn_casscf}
P.~E.~M. Siegbahn, J.~Alml\"{o}f, A.~Heiberg, and B.~O. Roos,
\newblock J. Chem. Phys. {\bf 74}, 2384 (1981).

\bibitem{Knowles_1985}
P.~J. Knowles and H.-J. Werner,
\newblock Chem. Phys. Lett. {\bf 115}, 259 (1985).

\bibitem{Werner_1985}
H.-J. Werner and P.~J. Knowles,
\newblock J. Chem. Phys. {\bf 82}, 5053 (1985).

\bibitem{molpro}
H.-J. {Werner} et~al.,
\newblock {MOLPRO, Version 2009.1, a Package of \emph{Ab initio} Programs,
  Cardiff University: Cardiff, United Kingdom, and University of Stuttgart:
  Stuttgart, Germany}.

\bibitem{Reiher_CsH}
G.~Moritz, A.~Wolf, and M.~Reiher,
\newblock J. Chem. Phys. {\bf 123}, 184105 (2005).

\bibitem{CASPT2_molpro}
H.-J. Werner,
\newblock Mol. Phys. {\bf 89}, 645 (1996).

\bibitem{Altmann}
S.~L. Altmann and P.~Herzig,
\newblock {\em Point-Group Theory Tables},
\newblock Oxford, 1994.

\bibitem{dmrg_ors}
O.~Legeza,
\newblock \textsc{QC-DMRG-Budapest}, a program for quantum chemical {DMRG}
  calculations. { \rm Copyright 2000--2013, HAS RISSPO Budapest}.

\bibitem{orbitalordering}
G.~Barcza, O.~Legeza, K.~H. Marti, and M.~Reiher,
\newblock Phys. Rev. A {\bf 83}, 012508 (2011).

\bibitem{legeza_dbss2}
O.~Legeza, J.~R\"oder, and B.~A. Hess,
\newblock Phys. Rev. B {\bf 67}, 125114 (2003).

\bibitem{nwchem}
M.~Valiev et~al.,
\newblock Comput.~Phys.~Commun. {\bf 181}, 1477 (2010).

\bibitem{nwchem_11}
H.~van Dam, W.~de~Jong, E.~Bylaska, N.~Govind, K.~Kowalski, T.~Straatsma, and
  M.~Valiev,
\newblock Rev. Comput. Mol. Sci. {\bf 1}, 888 (2011).

\bibitem{nwchem_web}
NWChem 6.1, {\tt http://www.nwchem-sw.org}.

\bibitem{tce_1}
S.~Hirata,
\newblock J.~Phys.~Chem.~A {\bf 107}, 9887 (2003).

\bibitem{tce_2}
S.~Hirata,
\newblock J.~Phys.~Chem. {\bf 121}, 51 (2004).

\bibitem{tce_3}
S.~Hirata, P.-D. Fan, A.~A. Auer, M.~Nooijen, and P.~Piecuch,
\newblock J.~Phys.~Chem. {\bf 121}, 12197 (2004).

\bibitem{tce_4}
K.~Kowalski, S.~Krishnamoorthy, R.~Olson, V.~Tipparaju, and E.~Apra,
\newblock Scalable implementations of accurate excited-state coupled cluster
  theories: {A}pplication of high-level methods to porphyrin-based systems; in:
  International conference for high performance computing, networking, storage
  and analysis,
\newblock in {\em International Conference for High Performance Computing,
  Networking, Storage and Analysis}, 2011.

\bibitem{MRCI_N2}
H.-J. Werner and P.~Knowles,
\newblock J. Chem. Phys. {\bf 94}, 1264 (1991).

\bibitem{Mitrushenkov_N2}
A.~O. Mitrushenkov, G.~Fano, F.~Ortolani, R.~Linguerri, and P.~Palmieri,
\newblock J. Chem. Phys. {\bf 115}, 6815 (2001).

\bibitem{Chan_N2}
G.~K.-L. Chan, M.~K\'{a}llay, and J.~Gauss,
\newblock J. Chem. Phys. {\bf 121}, 6110 (2004).

\bibitem{Shepard_2006}
R.~Shepard and M.~Minkoff,
\newblock Int. J. Quantum Chem. {\bf 106}, 3190 (2006).

\bibitem{Paldus_N2}
X.~Li and J.~Paldus,
\newblock J. Chem. Phys. {\bf 129}, 054104 (2008).

\bibitem{Shepard_2008}
R.~Shepard, G.~S. Kedziora, H.~Lischka, I.~Shavitt, T.~M\"{u}ller, P.~G.
  Szalay, M.~K\'{a}llay, and M.~Seth,
\newblock Chem. Phys. {\bf 349}, 37 (2008).

\bibitem{MRCC_N2_2009}
A.~Engels-Putzka and M.~Hanrath,
\newblock Mol. Phys. {\bf 107}, 143 (2009).

\bibitem{Piecuch_N2}
P.~Piecuch, S.~A. Kucharski, and K.~Kowalski,
\newblock Chem. Phys. Lett. {\bf 344}, 176 (2001).

\bibitem{CASPT2_N2}
T.~Fang, J.~Shen, and S.~Li,
\newblock J. Chem. Phys. {\bf 128}, 224107 (2008).

\bibitem{Jankowski_F2}
K.~Jankowski, R.~Becherer, P.~Scharf, H.~Schiffer, and R.~Ahlrichs,
\newblock J. Chem. Phys. {\bf 82}, 1413 (1985).

\bibitem{Ahlrichs_bs}
R.~Ahlrichs, P.~Scharf, and K.~Jankowski,
\newblock Chem. Phys. {\bf 98}, 381 (1985).

\bibitem{Paldus_F2}
X.~Li and J.~Paldus,
\newblock J. Chem. Phys. {\bf 108}, 637 (1998).

\bibitem{Piecuch_F2}
K.~Kowalski and P.~Piecuch,
\newblock Chem. Phys. Lett. {\bf 344}, 165 (2001).

\bibitem{Ivanov_F2}
V.~V. Ivanov, L.~Adamowicz, and D.~I. Lyakh,
\newblock Int. J. Quantum Chem. {\bf 106}, 2875 (2006).

\bibitem{Monika_F2}
M.~Musia{\l} and R.~J. Bartlett,
\newblock J. Chem. Phys. {\bf 122}, 224102 (2005).

\bibitem{Laidig_F2_N2}
W.~D. Laidig, P.~Saxe, and R.~J. Bartlett,
\newblock J. Chem. Phys. {\bf 86}, 887 (1987).

\bibitem{Carnell_CsH}
M.~Carnell, S.~D. Peyerimhoff, and B.~A. Hess,
\newblock Z. Phys. D. At., Mol. Clusters {\bf 333}, 317 (1989).

\bibitem{Zrafi_2006a}
W.~Zrafi, B.~Oujia, H.~Berriche, and F.~Gadea,
\newblock THEOCHEM {\bf 777}, 87 (2006).

\bibitem{Zrafi_2006b}
W.~Zrafi, N.~Khelif, B.~Oujia, and F.~X. Gadea,
\newblock J. Phys. B: At. Mol. Opt. Phys. {\bf 39}, 3815 (2006).

\bibitem{Yan_CsH}
L.~Yan, Y.~Qu, C.~Liu, J.~Wang, and R.~J. Buenker,
\newblock J. Chem. Phys. {\bf 136}, 124304 (2012).

\end{thebibliography}

\newpage
\begin{appendix}
\begin{center}
{\LARGE\bf Supporting Information}
\end{center}
%%%%%%%%%%%%%%%%%%%%%%%%%%%%%%%%%%%%%%%%%%%%%%%%%%%%%%%%%%%%%%%%%%%%%%%%%%%%%%%%%%%%%%%%%%%%%%%%%%%%%%%%%%%%%%%%%%%%%%%%%%
\section{DMRG calculations}\label{SI}
\subsection{Dissociation of N$_2$}
All electronic energies determined for different DMRG parameter sets and various interatomic
distances are summarized in Table \ref{tab:energiesn2}. We have employed the DBSS procedure
with a maximum number of renormalized active system states set to 1024 and a quantum
information loss of $10^{-5}$, while the minimum number of renormalized active system states
was varied as given in the Table. Note that for all calculations mentioned in Table \ref{tab:energiesn2} the maximum number of $m=1024$ was accepted for at most
one or two microiteration steps, and thus only a small number of renormalized active system
states is required to describe the system accurately.

In addition, we have performed some test calculations to check whether convergence has been
reached in cases where the energy difference between different converged DMRG calculations was slightly larger than 1.5 mHartree. In these
calculations, the initialization was performed with a minimum number of 1024 renormalized
active system states, while the remaining sweeps were performed with at least 512
renormalized active system states. Since the electronic energies differ by less than 0.3 mHartree
and similar entanglement diagrams have been obtained, the DMRG(10,46)[512,1024,$10^{-5}$] 
calculations are taken as reference. We should note that all DMRG calculations are converged higher
than 0.3 mHartree when comparing two subsequent parameter sets since the maximum number of renormalized active system states was
maintained for at most 2 microiteration steps, while the quantum information loss was two orders of magnitude lower than the chosen value of 10$^{-5}$ in all microiteration steps. 

\begin{table}[H]
\caption{Energies for N$_{2}$ in Hartree atomic units for DMRG(10,46)[$m_{\rm min}$,1024,$10^{-5}$] calculations for different interatomic distances $d_{\rm NN}$.}\label{tab:energiesn2}
{\footnotesize
\begin{center}
\begin{tabular}{lcccccc}\hline \hline
 Method		& \multicolumn{6}{c}{$E$/Hartree} \\ \cline{2-7}
                   & $d= 1.12$~\AA{}      &  $d=1.69$~\AA{}   & $d=2.12$~\AA{} &  $d=2.22$~\AA{}      & $d=2.33$~\AA{}  & $d=3.175$~\AA{} \\\hline
$m_{\rm min}=128$&$-$109.227 303	& $-$108.991 383		& $-$108.903 792		& $-$108.896 937		&$-$108.891 132		& $-$109.873 954\\
$m_{\rm min}=256$&$-$109.229 113	& $-$108.993 956		& $-$108.911 230		& $-$108.903 443		&$-$108.898 598		& $-$109.880 125\\
$m_{\rm min}=512$&$-$109.229 813	& $-$108.995 701		& $-$108.912 849		& $-$108.905 589		&$-$109.900 737		& $-$109.881 603\\
$m_{\rm min}=1024\backslash512$		&$-$109.229 847	&	&$-$108.913 120	&	&	$-$108.900 897	&
\\ \hline
\hline
\end{tabular}
\end{center}
}
\end{table}

\subsection{Dissociation of F$_2$}
The DMRG calculations for F$_2$ have been performed in a similar way as discussed
for the N$_2$ molecule and are summarized in Table \ref{tab:energiesf2}.
However, the accuracy of the initialization procedure (CI-DEAS)
severely determines convergence behavior. In order to avoid local minima, the number of
renormalized active system states had to be set to at least 1024 for the initialization steps
(these calculations are labeled as DMRG(14,32)[$m_{\rm start},m_{\rm min}$,$m_{\rm max}$,$10^{-5}$] in Table \ref{tab:energiesf2}). With such a large value for $m_{\rm start}$, convergence
with respect to $m$ can
be already obtained for a small number of renormalized active system states.

\begin{table}[H]
\small
\caption{Energies for F$_{2}$ in Hartree atomic units for DMRG(14,32)[($m_{\rm start},$)$m_{\rm min}$,$m_{\rm max}$,$10^{-5}$] calculations for different interatomic distances $d_{\rm FF}$.
If not mentioned otherwise, $m_{\rm start}$ has been set equal to $m_{\rm min}$.}\label{tab:energiesf2}
{
\begin{center}
\begin{tabular}{lcccc}\hline \hline
 				& \multicolumn{4}{c}{$E$/Hartree} \\ \cline{2-5}
Method 						& $d=1.41$ \AA &	$d=2.00$\AA	& $d=2.53$ \AA& $d=3.50$ \AA	\\ \hline
DMRG(14,32)[128,1024,10$^{-5}$]			&$-$198.936 599&$-$	&$-$198.878 361	&$-$198.876 518 \\
DMRG(14,32)[256,1024,10$^{-5}$]			&$-$198.937 016&$-$	&$-$198.895 701	&$-$198.889 183 \\
DMRG(14,32)[1024,256,1024,10$^{-5}$]&$-$198.968 768	&$-$199.930 466 & 	$-$						&		$-$				  \\
DMRG(14,32)[512,1024,10$^{-5}$]			&$-$		&$-$ &$-$198.908 453 	&$-$198.902 580 \\
DMRG(14,32)[1024,512,1024,10$^{-5}$]&$-$198.968 889	&$-$199.931 106 &$-$198.908 816	&$-$198.905 631 \\
DMRG(14,32)[1024,1500,10$^{-5}$]		&				$-$	&$-$		&$-$198.911 267	&$-$198.906 415
\\
DMRG(14,32)[1024,2048,10$^{-5}$]		&				$-$	&$-$		&$-$198.911 450   &
\\ \hline
\hline
\end{tabular}
\end{center}
}
\end{table}

\subsection{Dissociation of CsH}
Table \ref{tab:energiescsh} summarizes all electronic energies obtained from
DMRG calculations for the dissociation of the CsH molecule. As aforementioned,
the maximum number of renormalized active system states was set to 1024 with a
quantum information loss of $10^{-5}$, while the minimum number of renormalized
active system states was increased as mentioned in the Table until convergence
with respect to $m$ was reached.
Again, the maximum number of renormalized active-system states was kept for at most one   microiteration step, while the quantum information loss was considerably smaller than the chosen threshold of 10$^{-5}$. Hence, increasing $m_{\rm min}$ would result in an energy decrease of considerably less than 1 mHartree as observed for the N$_2$ molecules. 

\begin{table}[H]
\caption{Energies for CsH in Hartree atomic units for DMRG(10,$y$)[$m_{\rm min}$,1024,$10^{-5}$] calculations for different interatomic distances $d_{\rm CsH}$.
For $d_{\rm CsH} = 2.5$\AA{} and $d_{\rm CsH} = 7.0$\AA{}, 51 orbitals are contained in
the active space, while for $d_{\rm CsH} = 5.5$\AA{} only 50 orbitals could be included
due to symmetry considerations.}\label{tab:energiescsh}
{
\begin{center}
\begin{tabular}{lccccccc}\hline \hline
 		& \multicolumn{4}{c}{$E$/Hartree} \\ \cline{2-5}
Method		& $d=2.5$\AA{} 	&$d=3.50$\AA{}&  $d=5.5$\AA{}      & $d=7.0$\AA{}      \\\hline
 \\\hline
 $m_{\rm min}=128$		&$-$7783.946 277	& $-$	& $-$7783.884 358		& $-$7784.900 746	       \\
 $m_{\rm min}=256$		&$-$7783.970 497	& $-$7783.949 555	& $-$7783.907 986		& $-$7784.905 747	       \\
 $m_{\rm min}=512$		&$-$7783.971 143	& $-$7783.950 064	& $-$7783.908 452		& $-$7784.906 309	       
\\ \hline
\hline
\end{tabular}
\end{center}
}
\end{table}

\subsection{Additional entanglement diagrams}

\begin{figure}[H]
\centering
\includegraphics[width=0.6\linewidth]{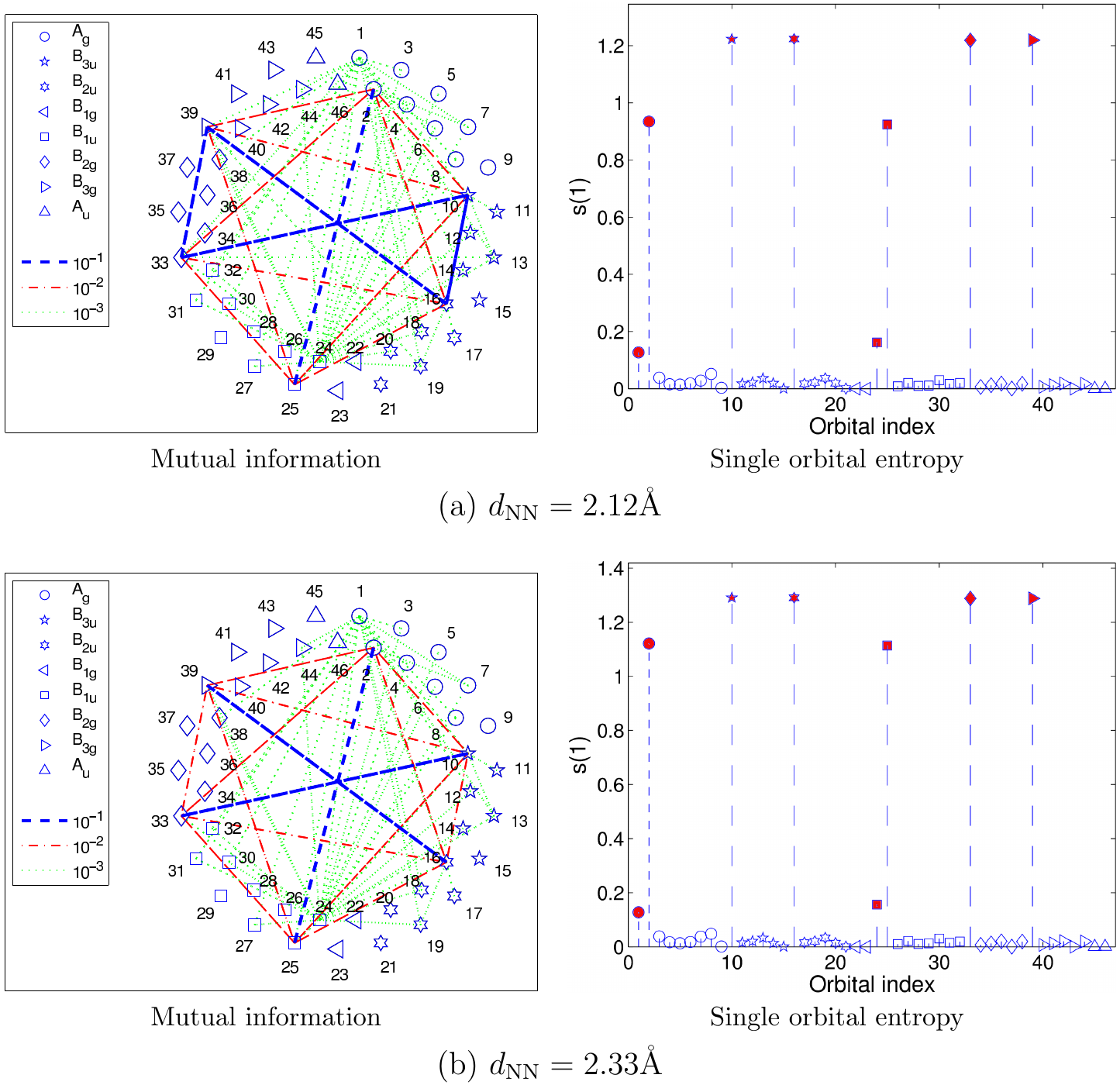}
\caption{
Mutual information and single orbital entropies $s(1)_i$ for DMRG(10,46)[512,1024,$10^{-5}$]
calculations for the N$_2$ molecule at different internuclear distances.
The orbitals are numbered and sorted according to their (CASSCF) natural occupation numbers.
Each orbital index in the $s(1)_i$ diagram indicates one molecular orbital and corresponds
to the same natural orbital as numbered in the mutual information plot.
Those orbitals which are included in the CAS(10,8)SCF calculations are marked in red in the
single-orbital entropy diagram.
The total quantum information sums up to 7.63, and 8.26 for increasing distances.
}\label{fig:n2-complement}
\end{figure}

\end{appendix}

\end{document}